\documentclass[12pt,draftcls,onecolumn]{IEEEtran}
\usepackage{graphicx}
\usepackage{amstext}
\usepackage{algorithm}
\usepackage{algorithmic}
\usepackage{mathrsfs}
\usepackage{amssymb}
\usepackage{amsmath}
\usepackage{bm}
\allowdisplaybreaks[4]
\usepackage{epstopdf}
\usepackage{multicol}
\usepackage{stfloats}
\usepackage{color}
\usepackage{enumerate}
\usepackage[colorlinks,
            linkcolor=black,
            anchorcolor=blue,
            citecolor=black
            ]{hyperref}
\usepackage{breqn}
\usepackage{bbm}
\usepackage{cite}
\usepackage{subfig}
\usepackage{caption}
\newtheorem{Lemma}{Lemma}
\newtheorem{Corollary}{Corollary}
\begin{document}
\title{\huge Random Caching Based Cooperative Transmission in Heterogeneous Wireless Networks}

\author{Wanli~Wen,~\IEEEmembership{Student~Member,~IEEE,}
        Ying~Cui,~\IEEEmembership{ Member,~IEEE,}
        Fu-Chun~Zheng,~\IEEEmembership{Senior Member,~IEEE,}
        Shi~Jin,~\IEEEmembership{ Member,~IEEE}
        and~Yanxiang~Jiang,~\IEEEmembership{ Member,~IEEE}
\thanks{W. Wen, F.-C. Zheng, S. Jin and Y. Jiang are with the National Mobile Communications
Research Laboratory, Southeast University, Nanjing 210093, China (e-mail:wenwanli@seu.edu.cn; fzheng@seu.edu.cn; jinshi@seu.edu.cn; yxjiang@seu.edu.cn). Y. Cui is with the Department of Electronic Engineering,
Shanghai Jiao Tong University, Shanghai 200240, China (e-mail: cuiying@sjtu.edu.cn).}}% <-this % stops a space
\maketitle
\vspace{-1cm}
% As a general rule, do not put math, special symbols or citations
% in the abstract or keywords.
\begin{abstract}
Base station cooperation in heterogeneous wireless networks (HetNets) is a promising approach to improve the network performance, but it also imposes a significant challenge on backhaul. On the other hand, caching at small base stations (SBSs) is considered as an efficient way to reduce backhaul load in HetNets. In this paper, we jointly consider SBS caching and cooperation in a downlink large-scale HetNet. We propose two SBS cooperative transmission schemes under random caching at SBSs with the caching distribution as a design parameter. Using tools from stochastic geometry and adopting appropriate integral transformations, we first derive a tractable expression for the successful transmission probability under each scheme. Then, under each scheme, we consider the successful transmission probability maximization by optimizing the caching distribution, {\color{black}which is a challenging optimization problem with a non-convex objective function.} By exploring optimality properties and using optimization techniques, under each scheme, we obtain a local optimal solution in the general case and global optimal solutions in some special cases. Compared with some existing caching designs in the literature, e.g., the most popular caching, the i.i.d. caching and the uniform caching, the optimal random caching under each scheme achieves better successful transmission probability performance. The analysis and optimization results provide valuable design insights for practical HetNets.
\end{abstract}
%
%% Note that keywords are not normally used for peerreview papers.

\IEEEpeerreviewmaketitle

\section{Introduction}
\IEEEPARstart{D}{ue} to the explosive growth of mobile data traffic, the demand for wireless communication services has been shifting from connection-oriented services such as traditional voice telephony and messaging to content-oriented services such as multimedia, social networking and smartphone applications. Recently, heterogeneous networks (HetNets) \cite{HeterogeneouscellularnetworksFromtheorytopractice} in which dense small base stations (SBSs), e.g., pico BSs and femto BSs, are deployed along with the existing macro base stations (MBSs) are considered as an attractive solution to meet the ever increasing mobile data traffic demand. In order to address the additional intercell interference caused by such deployment, BS cooperation in HetNets has been proposed as one of the solutions to effectively mitigate the interference at mobile stations, but it also imposes on significant challenge on the backhaul.

{BS joint transmission, consisting of non-coherent \cite{CoordinatedmultipointConceptsperformanceandfieldtrialresults, Fivedisruptivetechnologydirectionsfor5G, CoordinatedmultipointtransmissionreceptiontechniquesforLTE, ATractableModelforNoncoherentJointTransmissionBaseStationCooperation, Stochasticgeometricanalysisofhandoffsinusercentriccooperativewirelessnetworks, 3, Analysisofnoncoherentjointtransmissioncooperationinheterogeneouscellularnetworks,  UserCentricCrossTierBaseStationClusteringandCooperation,  Energyefficientbasestationcooperationindownlinkheterogeneouscellularnetworks, CoverageAnalysisforCoMPinTwoTierHetNetsWithNonuniformlyDeployedFemtocells} and coherent \cite{AStochasticGeometryFrameworkforAnalyzingPairwiseCooperativeCellularNetworks} joint transmissions, is one of the much studied BS cooperation schemes.} In non-coherent joint transmission, BSs cooperate by jointly transmitting the same data to a user without prior phase alignment. In contrast, in coherent joint transmission, BSs jointly transmit the same data to a user {with prior phase alignment,} assuming that stringent synchronization can be done and perfect channel state information (CSI) is available at all cooperative BSs. If these strict requirements can be satisfied, coherent joint transmission achieves better performance. Otherwise, non-coherent joint transmission is more preferable, especially, in lightly-loaded scenarios \cite{ATractableModelforNoncoherentJointTransmissionBaseStationCooperation}. Due to its low complexity and requirement, BS non-coherent joint transmission in large-scale HetNets has been widely considered and extensively studied using some effective tools from stochastic geometry \cite{3, Analysisofnoncoherentjointtransmissioncooperationinheterogeneouscellularnetworks, UserCentricCrossTierBaseStationClusteringandCooperation, Energyefficientbasestationcooperationindownlinkheterogeneouscellularnetworks,  CoverageAnalysisforCoMPinTwoTierHetNetsWithNonuniformlyDeployedFemtocells}. The number of BSs jointly serving a user located at the origin (referred to as the typical user) is fixed in \cite{3}, and is variable in \cite{Analysisofnoncoherentjointtransmissioncooperationinheterogeneouscellularnetworks,  UserCentricCrossTierBaseStationClusteringandCooperation, Energyefficientbasestationcooperationindownlinkheterogeneouscellularnetworks}. In particular, in \cite{3}, the BSs with the strongest average received powers at the typical user form the BS cooperation set. In \cite{Analysisofnoncoherentjointtransmissioncooperationinheterogeneouscellularnetworks} and \cite{UserCentricCrossTierBaseStationClusteringandCooperation}, the BSs with instantaneous received power at the typical user above some thresholds (one for each tier) form the BS cooperation set, and the optimization of the thresholds is considered in \cite{UserCentricCrossTierBaseStationClusteringandCooperation}. In \cite{Energyefficientbasestationcooperationindownlinkheterogeneouscellularnetworks}, the BSs within a circle of a tunable radius centered at the typical user jointly serve the typical user, and the optimization of the radius is considered. In \cite{CoverageAnalysisforCoMPinTwoTierHetNetsWithNonuniformlyDeployedFemtocells}, the authors consider a user located at macro cell edge and propose a cooperation scheme to serve the user by its geographically nearest MBS and SBS, under certain conditions. Note that, the non-coherent joint transmission for HetNets in \cite{3,Analysisofnoncoherentjointtransmissioncooperationinheterogeneouscellularnetworks,  Energyefficientbasestationcooperationindownlinkheterogeneouscellularnetworks, UserCentricCrossTierBaseStationClusteringandCooperation,  CoverageAnalysisforCoMPinTwoTierHetNetsWithNonuniformlyDeployedFemtocells} imposes a significant challenge on the backhaul.

In practice, the backhaul has increasingly become a bottleneck, which limits the potential of BS joint transmission in HetNets. In order to alleviate the backhaul load caused by the BS joint transmission, the authors in \cite{SpatiotemporalCooperationinHeterogeneousCellularNetworks} purpose a BS silencing scheme in large-scale HetNets, where the typical user is served by its nearest BS and the nearby BSs keep silent to facilitate the transmission. Reference \cite{SpatiotemporalCooperationinHeterogeneousCellularNetworks} further shows that compared with joint transmission, BS silencing yields a lower complexity and a lighter backhaul load, at the cost of coverage probability.

Caching at SBSs has been proposed as a promising approach for remarkably reducing backhaul load by prefetching popular files into storages at SBSs {in large-scale small cell networks or HetNets} \cite{Cache-enabledsmallcellnetworksmodelingandtradeoffs, EdgeCachingforCoverageandCapacityaidedHeterogeneousNetworks, CacheenabledheterogeneouscellularnetworksComparisonandtradeoffs, CachinginWirelessSmallCellNetworksAStorageBandwidthTradeoff, Tamoorulhassan2015Modeling, Bharath2015Caching, ALearningBasedApproachtoCachinginHeterogenousSmallCellNetworks}. In \cite{Cache-enabledsmallcellnetworksmodelingandtradeoffs, EdgeCachingforCoverageandCapacityaidedHeterogeneousNetworks, CacheenabledheterogeneouscellularnetworksComparisonandtradeoffs, CachinginWirelessSmallCellNetworksAStorageBandwidthTradeoff}, the authors consider caching the most popular files at each SBS, which we in this paper refer to as ``most popular caching (MPC)". In \cite{Tamoorulhassan2015Modeling}, the authors consider random caching with uniform distribution at SBSs, assuming that file requests follow a uniform distribution, which we call ``uniform caching (UC)". In \cite{Bharath2015Caching} and \cite{ ALearningBasedApproachtoCachinginHeterogenousSmallCellNetworks}, the authors consider random caching with files being stored at each SBS in an i.i.d. manner, which we refer to as ``i.i.d. caching (IIDC)". The MPC scheme considered in \cite{Cache-enabledsmallcellnetworksmodelingandtradeoffs, EdgeCachingforCoverageandCapacityaidedHeterogeneousNetworks, CacheenabledheterogeneouscellularnetworksComparisonandtradeoffs, CachinginWirelessSmallCellNetworksAStorageBandwidthTradeoff} does not provide any
spatial file diversity. In contrast, the caching designs in \cite{Tamoorulhassan2015Modeling, Bharath2015Caching, ALearningBasedApproachtoCachinginHeterogenousSmallCellNetworks} can provide file diversity. However, the UC scheme in \cite{Tamoorulhassan2015Modeling} only provides caching probabilities of files and does not specify how multiple different files can be efficiently stored at each SBS based on these probabilities. The IIDC scheme in \cite{Bharath2015Caching} and \cite{ALearningBasedApproachtoCachinginHeterogenousSmallCellNetworks} may waste storage resources, as multiple copies of the same file may be stored at one SBS. Hence, the caching designs in \cite{Cache-enabledsmallcellnetworksmodelingandtradeoffs, EdgeCachingforCoverageandCapacityaidedHeterogeneousNetworks, CacheenabledheterogeneouscellularnetworksComparisonandtradeoffs, CachinginWirelessSmallCellNetworksAStorageBandwidthTradeoff, Tamoorulhassan2015Modeling, Bharath2015Caching, ALearningBasedApproachtoCachinginHeterogenousSmallCellNetworks} may not yield the best network performance.

As a result, some other works have considered optimal caching designs in large-scale small cell networks or HetNets \cite{OptimalGeographicCachingInCellularNetworks, OptimalContentPlacementforOffloadinginCacheenabledHeterogeneousWirelessNetworks, CuiCacheWirelessSignleTier, AnalysisandOptimizationofCachingandMulticastinginLargeScaleCacheEnabledHeterogeneousWirelessNetworks,  CachingPlacementinStochasticWirelessCachingHelperNetworksChannelSelectionDiversityviaCaching}. {\color{black}In \cite{OptimalGeographicCachingInCellularNetworks} and \cite{OptimalContentPlacementforOffloadinginCacheenabledHeterogeneousWirelessNetworks}, the authors consider random caching at SBSs, and analyze and optimize the cache hit probability (i.e., the probability that a randomly requested file from the typical user is stored at its serving BS) \cite{OptimalGeographicCachingInCellularNetworks} and successful offloading probability (i.e., the probability that the typical user is associated with the SBS tier and its downlink signal-to-interference-plus-noise ratio is larger than a threshold) \cite{OptimalContentPlacementforOffloadinginCacheenabledHeterogeneousWirelessNetworks}.} In \cite{CuiCacheWirelessSignleTier}, the authors consider random caching and multicasting at SBSs in a large-scale  small cell network, and analyze and optimize the successful transmission probability (i.e., the probability that a randomly requested file from the typical user can be successfully received). In \cite{AnalysisandOptimizationofCachingandMulticastinginLargeScaleCacheEnabledHeterogeneousWirelessNetworks}, the authors propose a hybrid caching design (consisting of identical caching in the MBS tier and random caching in the SBS tier) and a corresponding multicasting design in a large-scale HetNet, and analyze and optimize the successful transmission probability. In \cite{CachingPlacementinStochasticWirelessCachingHelperNetworksChannelSelectionDiversityviaCaching}, the authors investigate how channel selection diversity affects the optimal random caching design. Note that, none of the above works \cite{Cache-enabledsmallcellnetworksmodelingandtradeoffs, EdgeCachingforCoverageandCapacityaidedHeterogeneousNetworks, CacheenabledheterogeneouscellularnetworksComparisonandtradeoffs, CachinginWirelessSmallCellNetworksAStorageBandwidthTradeoff, Tamoorulhassan2015Modeling, Bharath2015Caching, ALearningBasedApproachtoCachinginHeterogenousSmallCellNetworks, OptimalGeographicCachingInCellularNetworks, OptimalContentPlacementforOffloadinginCacheenabledHeterogeneousWirelessNetworks, CuiCacheWirelessSignleTier, AnalysisandOptimizationofCachingandMulticastinginLargeScaleCacheEnabledHeterogeneousWirelessNetworks,  CachingPlacementinStochasticWirelessCachingHelperNetworksChannelSelectionDiversityviaCaching} has considered SBS cooperation.

In \cite{CooperativeTransmissionviaCachingHelpers, DistributedCachingandSmallCellCooperationforFastContentDelivery, CooperativeCachingandTransmissionDesigninClusterCentricSmallCellNetworks}, the authors jointly consider SBS caching and cooperation in large-scale small cell networks or HetNets. Specifically, in \cite{CooperativeTransmissionviaCachingHelpers}, the SBSs storing the requested file and within a circle of a certain radius centered at the typical user jointly serve the typical user. In \cite{DistributedCachingandSmallCellCooperationforFastContentDelivery}, a certain number of SBSs (i.e., with the same distances to the typical user) storing the requested file jointly serve the typical user. The optimal caching designs in \cite{CooperativeTransmissionviaCachingHelpers, DistributedCachingandSmallCellCooperationforFastContentDelivery} are obtained by maximizing the successful transmission probability. In \cite{CooperativeCachingandTransmissionDesigninClusterCentricSmallCellNetworks}, the authors propose a partion-based combined caching design, where a certain number of SBSs storing the requested file jointly serve the typical user. However, in \cite{CooperativeTransmissionviaCachingHelpers}, the cache size of each SBS is assumed to be one, and thus, the impact of the cache size in practical networks cannot be addressed; in \cite{DistributedCachingandSmallCellCooperationforFastContentDelivery}, the distances between the cooperative SBSs and the typical user are fixed and identical, and thus, the stochastic nature of geographic locations of cooperative SBSs cannot be all captured; the combined caching design in \cite{CooperativeCachingandTransmissionDesigninClusterCentricSmallCellNetworks} cannot reflect the popularity differences among some files, and hence may not yield the best possible performance.

{Therefore, further studies are required to reveal how SBS caching and cooperation can jointly and optimally affect the network performance of HetNets. In this paper, we address these issues. Our main contributions are summarized below.}
\begin{itemize}
  \item We propose two SBS cooperative transmission schemes under random caching at SBSs with the caching distribution as a design parameter. Specifically, the first scheme adopts non-coherent joint transmission, and the second scheme effectively combines non-coherent joint transmission and BS silencing.
  \item We analyze the successful transmission probability. SBS cooperation and random caching make the analysis very challenging. By using tools from stochastic geometry and adopting appropriate integral transformations, under each scheme, we derive a tractable expression for the successful transmission probability.
  \item We consider the successful transmission probability maximization by optimizing the caching distribution, which is a very challenging optimization problem with a non-convex objective function. By exploring optimality properties and using optimization techniques, we obtain a local optimal solution in the general case and global optimal solutions in some special cases, under each~scheme.
  \item {\color{black}By numerical results, we show that the optimal caching distributions are influenced by multiple system parameters jointly, such as the file popularity, the cache size and the number of cooperative SBSs, etc.} In addition, we also show that under each scheme, the optimal caching design achieves a significant gain in the successful transmission probability over some existing caching designs in the literature, e.g., MPC, IIDC and UC.
%  \item By numerical results, we show that under each scheme, the optimal caching design achieves a significant gain in the successful transmission probability over MPC, IIDC and UC. In addition, the first scheme outperforms the second one when the target bit rate is small, and underperforms the second one when the target bit rate is large.
\end{itemize}
\section{System Model}
\begin{figure*}
  \centering%\textwidth
  \includegraphics[width=6in]{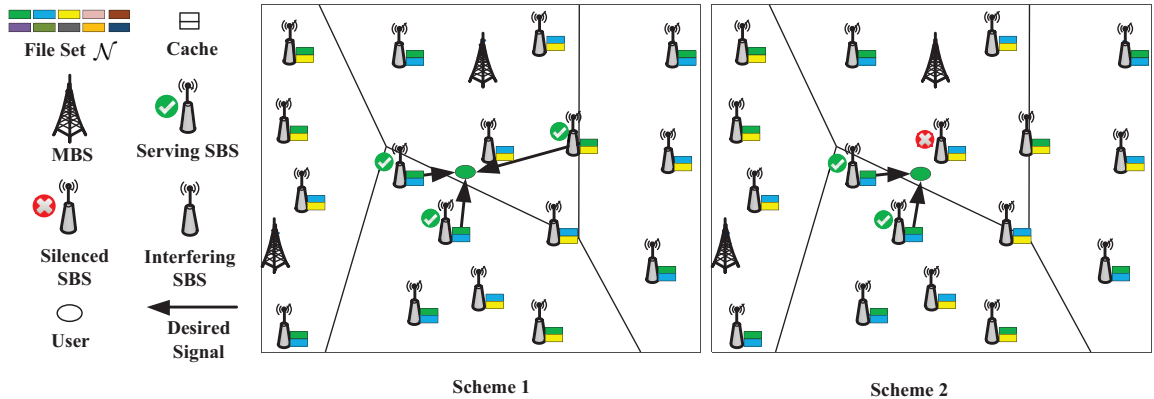}
  \caption{Illustration of SBS cooperation under Scheme~1 and Scheme~2. The MBS tier corresponds to a Voronoi tessellation, determined by the locations of all the MBSs. The color of the typical user corresponds to the file it requests. $|\mathcal{C}_{1,n}| = |\mathcal{C}_2| = 3$, $|\mathcal{C}_{2,n}|=2$, $M=2$ and $N=10$.}\label{sysmod}
\end{figure*}
We consider a downlink two-tier HetNet where a tier of MBSs are overlaid with a tier of much denser SBSs, as shown in Fig. \ref{sysmod}. The locations of the SBSs and MBSs are spatially distributed as two independent homogeneous Poisson point processes (PPPs) ${\Phi_{\textsf{s}}}$ and ${\Phi_{\textsf{m}}}$ with densities ${\lambda _{\textsf{s}}}$ and ${\lambda_{\textsf{m}}}$ (${\lambda _{\textsf{s}}} > {\lambda _{\textsf{m}}}$), respectively. For the ease of illustration, we use subscripts $\textsf{s}$ and $\textsf{m}$ to distinguish the SBS tier and the MBS tier. The transmission powers at each SBS and MBS are ${P_{\textsf{s}}}$ and $P_{\textsf{m}}$ (${P_{\textsf{s}}} < {P_{\textsf{m}}}$), respectively. We assume that users are also distributed according to an independent homogeneous PPP and focus on studying a typical user $u_0$ located at the origin. We adopt universal frequency reuse for each BS over the entire frequency band. Each BS equally divides its total bandwidth to serve all the users associated with it. The available bandwidths of each SBS and MBS for $u_0$ are represented by $W_{\textsf{s}}$ and $W_{\textsf{m}}$ ($W_{\textsf{s}}>W_{\textsf{m}}$), respectively.\footnote{In the traditional connection-based HetNets, a user is associated with the specific BS, which provides the maximum received signal strength \cite{ModelingandAnalysisofKTierDownlinkHeterogeneousCellularNetworks,LoadAwareModelingandAnalysisofHeterogeneousCellularNetworks}. The transmit power disparity of SBSs and MBSs (${P_{\textsf{s}}} < {P_{\textsf{m}}}$) will lead to the association of more users with the MBS than the SBS. Thus, the available bandwidth of the SBS for $u_0$ is in general less than that of the MBS.} {The typical user $u_0$ and all BSs are equipped with a single antenna.}\footnote{Note that, the analytical framework developed in this paper can be extended to the multi-antenna scenario.} {Due to large-scale path-loss, a transmitted signal from an MBS (SBS) with distance $r$ is attenuated by a factor $r^{-\alpha_{\textsf{m}}}$ ($r^{-\alpha_{\textsf{s}}}$), where $\alpha_{\textsf{m}}>2$ ($\alpha_{\textsf{s}}>2)$ is the path-loss exponent for MBSs (SBSs). For small-scale fading, we assume Rayleigh fading channels.}

Denote by ${\mathcal{N}} = \{ 1,2, \cdots ,N\} $ the set of $N\in \mathbbm{N}$ files in the HetNet, where $\mathbbm{N}$ denotes the set of natural numbers. For ease of analysis, we assume that all files have the same size. File $n \in {\mathcal{N}}$ is requested with probability ${a_n} \in (0,1)$, where $\sum\nolimits_{n = 1}^N {{a_n}}  = 1$. In addition, without loss of generality (w.l.o.g), we assume that ${a_1} > {a_2} >  \cdots > {a_N}$.

We assume that each MBS is equipped with no cache but is connected to the core network via optical fibers with high capacity. Thus, each MBS can retrieve all files from the core network. In this paper, we ignore file downloading costs at MBS. Each SBS is equipped with a cache of size $M$ (in files), where $M \le N$, and can serve any files stored locally. To provide spatial file diversity (which can improve performance of dense wireless networks) \cite{CuiCacheWirelessSignleTier}, we adopt a random caching scheme at SBSs. In particular, each SBS stores $M$ different files out of all $N$ files in ${{\mathcal{N}}}$ with a certain probability. Let $T_n$ denote the probability of file $n$ being stored at an SBS. Denote $\mathbf{T}\triangleq\left(T_n\right)_{n\in \mathcal{N}}$, which is termed as the ``caching distribution". Then, we have \cite{CuiCacheWirelessSignleTier}:
\begin{eqnarray}\label{eqsysmod1}
0\le T_n \le 1,\label{Tncons1}\\
\sum\nolimits_{n\in \mathcal{N}}T_n=M.\label{Tncons2}
\end{eqnarray} Let $\Phi_{\textsf{s},n}$ denote the set of the SBSs which store file $n$. Note that $\Phi_{\textsf{s},n}=\emptyset$ if $T_n=0$, and $\Phi_{\textsf{s}}\triangleq\bigcup{_{n \in \mathcal{N}}} {{\Phi_{\textsf{s},n}}}$. By \cite{Haenggi2012Stochastic}, we know that $\Phi_{\textsf{s},n}$ is also a homogeneous PPP with density $\lambda_{\textsf{s}} T_n$.

In the following illustration, we suppose $u_0$ requests file $n$. Assume all MBSs are active. First, we introduce some notations. According to the distance between each SBS and $u_0$, let $\mathcal{C}_{1,n}$ denote the set of $u_0$'s $K$ nearest SBSs in $\Phi_{\textsf{s},n}$, and let $\mathcal{C}_2$ denote the set of $u_0$'s $K$ nearest SBSs in $\Phi_{\textsf{s}}$. Denote $\mathcal{C}_{2,n}\triangleq {\mathcal{C}_2\bigcap\Phi_{\textsf{s},n}}$ and $\mathcal{C}_{2,-n}\triangleq \mathcal{C}_2\setminus \mathcal{C}_{2,n}$. Now, we propose two cooperative transmission schemes.
\begin{itemize}
  \item Scheme 1: If $T_n=0$ (i.e., file $n$ is not stored at any SBS), $u_0$ is served by its nearest MBS. If $T_n>0$ (i.e., file $n$ is stored at some SBSs), all SBSs in $\mathcal{C}_{1,n}$ ($\left|\mathcal{C}_{1,n}\right|=K$) jointly transmit file $n$ to $u_0$. In both cases, all SBSs in $\Phi_{\textsf{s}}\setminus \mathcal{C}_{1,n}$ are assumed to be active to serve other~users.
  \item Scheme 2: {If $\mathcal{C}_{2,n}= \emptyset$, the nearest MBS will serve $u_0$, and all SBSs in $\Phi_{\textsf{s}}$ are assumed to be active to serve other users. If $\mathcal{C}_{2,n}\neq \emptyset$, all SBSs in $\mathcal{C}_{2,n}\subseteq \mathcal{C}_{2}$ jointly transmit file $n$ to $u_0$, and all SBSs in ${\cal C}_{2,-n}$ are silenced to facilitate the transmission of file $n$ to $u_0$, and all SBSs in $\Phi_{\textsf{s}}\setminus \mathcal{C}_2$ are assumed to be active to serve other users.}
\end{itemize}

Fig. \ref{sysmod} illustrates the SBS cooperation scenario under the two schemes. Note that, under each scheme, we refer to an SBS in $\mathcal{C}_{1,n}$ or $\mathcal{C}_{2,n}$ as a serving SBS and an SBS in $\Phi_{\textsf{s}}\setminus \mathcal{C}_{1,n}$ or $\Phi_{\textsf{s}}\setminus \mathcal{C}_{2}$ as an interfering SBS. Similarly, when $u_0$ is served by its nearest MBS, we refer to the nearest MBS as the serving MBS and other MBSs as interfering MBSs. Assume that CSI is not known at any BS. Thus, we cannot adopt prior phase correction at cooperative SBSs, and will now instead consider non-coherent joint transmission \cite{3}.
\newtheorem{Remark}{Remark}
\begin{Remark}[Comparison between Scheme 1 and Scheme 2]\label{remarkcompasch1sch2}
When cooperative SBSs in $\mathcal{C}_{1,n}$ (under Scheme~1) and $\mathcal{C}_{2,n}$ (under Scheme~2) jointly transmit the same data to $u_0$, we have the following statements.
\begin{enumerate}
    \item The number of serving SBSs under Scheme~2 (i.e., $|\mathcal{C}_{2,n}|$) is a random variable with the mean of $T_nK$, while the number of serving SBSs under Scheme~1 (i.e., $K$) is fixed. Assuming $T_n<1$, the average number of serving SBSs under Scheme~2 is always smaller than that under Scheme~1, and the average received signal power under Scheme~2 is weaker than that under Scheme~1.
    \item The numbers of interfering SBSs under the two schemes are the same. The average interference power under Scheme~1 is stronger than that under Scheme~2.
    \item If $T_n=1$ for all $n=1,\cdots,M$ and $T_n=0$ for all $n=M+1,M+2,\cdots,N$ (i.e., the MPC design is adopted at all SBSs), Scheme 1 and Scheme 2 become the same~scheme.
\end{enumerate}
\end{Remark}

We consider an interference-limited network and neglect the background thermal noise \cite{CooperativeCachingandTransmissionDesigninClusterCentricSmallCellNetworks}. We now derive the instantaneous received signal-to-interference ratio (SIR) at ${u_0}$. Let ${h_{x,l}}$ and ${r_{x,l}}$ denote the {\color{black}fading and distance between BS ${l}$ in tier $x\in\{\textsf{s},\textsf{m}\}$} and ${u_0}$, respectively. {\color{black}Let $l_{\textsf{m}}$ and $l_{\textsf{s}}$ denote the indexes of the serving MBS and SBS of $u_0$, respectively.} Thus, under each scheme, if $u_0$ is served by its serving MBS, the SIR at $u_0$ is given by
\begin{equation}\label{eqDefineSIRGammam}
\gamma _{\textsf{m}} = \frac{{{P_{\textsf{m}}}{|h_{\textsf{m},l_{\textsf{m}}}|^2}r_{\textsf{m},l_{\textsf{m}}}^{ - {\alpha_{\textsf{m}}}}}}{{\sum\limits_{{l} \in \Phi_{\textsf{s}}} {{P_{\textsf{s}}}{{\left| {{h_{\textsf{s},l}}} \right|}^2}r_{\textsf{s},l}^{ - {\alpha_{\textsf{s}}}}} + \sum\limits_{{l} \in \Phi_{\textsf{m}}\backslash \{l_{\textsf{m}}\}} {{P_{\textsf{m}}}{{\left| {{h_{\textsf{m},l}}} \right|}^2}r_{\textsf{m},l}^{ - {\alpha_{\textsf{m}}}}}}}.
\end{equation}Otherwise, $u_0$ non-coherently combines desired signals from serving SBSs by accumulating their amplitudes \cite{GPPCoMP}, and the SIRs at ${u_0}$ under Scheme 1 and Scheme 2 are given by
\setlength{\arraycolsep}{0.0em}\begin{eqnarray}
\gamma_{\textsf{s}_1,n} &=& \frac{{\left| {\sum\limits_{{l_{\textsf{s}}} \in {\mathcal{C}_{1,n}}} {\sqrt {{P_{\textsf{s}}}} {h_{\textsf{s},l_{\textsf{s}}}}r_{\textsf{s},l_{\textsf{s}}}^{ - {\alpha_{\textsf{s}}}/2}} } \right|^2}}{{\sum\limits_{{l} \in \Phi_{\textsf{s}}\backslash {\mathcal{C}_{1,n}}} {{P_{\textsf{s}}}{{\left| {{h_{\textsf{s},l}}} \right|}^2}r_{\textsf{s},l}^{ - {\alpha_{\textsf{s}}}}} + {\sum\limits_{{l} \in \Phi_{\textsf{m}}} {{P_{\textsf{m}}}{{\left| {{h_{\textsf{m},l}}} \right|}^2}r_{\textsf{m},l}^{ - {\alpha_{\textsf{m}}}}}}}},\label{eqgammas1n}\\
{\gamma_{\textsf{s}_2,n}} &=&
\frac{{{{\left| {\sum\limits_{{l_{\textsf{s}}} \in {{\cal C}_{2,n}}} {\sqrt {{P_{\textsf{s}}}} {h_{\textsf{s},l_{\textsf{s}}}}r_{\textsf{s},l_{\textsf{s}}}^{ - {\alpha_{\textsf{s}}}/2}} } \right|}^2}}}{{{\sum \limits_{{l} \in {\Phi_{\textsf{s}}}\backslash {{\cal C}_{2}}}}{P_{\textsf{s}}}{{\left| {{h_{\textsf{s},l}}} \right|}^2}r_{\textsf{s},l}^{ - {\alpha_{\textsf{s}}}} + \sum\limits_{{l} \in {\Phi_{\textsf{m}}}} {{P_{\textsf{m}}}{{\left| {{h_{\textsf{m},l}}} \right|}^2}r_{\textsf{m},l}^{ - {\alpha_{\textsf{m}}}}}}}.\label{eqgammas2n}
\end{eqnarray}\setlength{\arraycolsep}{5pt}

In this paper, we employ the successful transmission probability (STP) \cite{AnalysisandOptimizationofCachingandMulticastinginLargeScaleCacheEnabledHeterogeneousWirelessNetworks} as the system performance metric. {\color{black}Each file is transmitted at a target bit rate $\tau$ (bps). $u_0$ successfully receives file $n$ if the channel capacity between the serving MBS or SBSs  and $u_0$ exceeds $\tau$. Let $\psi_{\mathrm{sch}_i}(\mathbf{T})$ denote the STP under Scheme $i$, $i=1,2$. Then we have:}
\setlength{\arraycolsep}{0.0em}\begin{eqnarray}\label{psi1Def}
\psi_{\mathrm{sch}_1}(\mathbf{T}) &=& \sum\nolimits_{n \in \mathcal{N}}{a_n}\left( \Pr \left[ \tau_{\textsf{m}} > \tau\right]\mathbbm{1}\left[T_n=0\right]+\Pr \left[ \tau_{\textsf{s}_1} > \tau \right]\mathbbm{1}\left[T_n>0\right]\right),\\
\psi_{\mathrm{sch}_2}(\mathbf{T}) &=& \sum\nolimits_{n \in \mathcal{N}}{a_n}\left(\Pr \left[ \tau_{\textsf{m}} > \tau,C_{2,n}=0\right]+ \sum\nolimits_{k=1}^{K}\Pr \left[ \tau_{\textsf{s}_2} > \tau ,C_{2,n}=k\right]\right),\label{psi2Def}
\end{eqnarray}\setlength{\arraycolsep}{5pt}where $\tau_{\textsf{m}}\triangleq {W_{\textsf{m}}{{\log }_2}\left( {1 + {\gamma _{\textsf{m}}}} \right) }$ represents the channel capacity between the serving MBS and $u_0$ under both schemes, $\tau_{\textsf{s}_1}\triangleq W_{\textsf{s}}{{\log }_2}\left( {1 + {\gamma _{\textsf{s}_1,n}}} \right)$ and $\tau_{\textsf{s}_2}\triangleq W_{\textsf{s}}{{\log }_2}\left( {1 + {\gamma _{\textsf{s}_2,n}}} \right)$ represent the channel capacity between the serving SBSs and $u_0$ under Scheme 1 and Scheme 2, respectively, $\mathbbm{1}[\bullet]$ denotes the indicator function, and $C_{2,n}\triangleq|\mathcal{C}_{2,n}|$ denotes the number of the SBSs storing file $n$ in $\mathcal{C}_{2,n}$.
\section{Analysis and Optimization of Performance under Scheme 1}
In this section, we first analyze the STP under Scheme 1 for a given caching distribution $\mathbf{T}$ of the random caching scheme. Then we maximize the STP by optimizing $\mathbf{T}$.
\subsection{Analysis of Successful Transmission Probability}
In this part, we analyze the STP ${\psi_{\mathrm{sch}_1}(\mathbf{T})}$ under Scheme 1 using tools from stochastic geometry. {When $u_0$ is served by its serving MBS, as in the traditional connection-based HetNets, $\Pr\left[ \tau_{\textsf{m}}>\tau\right]$ can be calculated using the method in \cite{Cache-enabledsmallcellnetworksmodelingandtradeoffs}. When $u_0$ is served by its serving SBSs, different from the traditional connection-based HetNets, there are three types of interferers, namely, i) all the other SBSs storing the desired file of $u_0$ besides its serving SBSs, ii) all the SBSs without the desired file of $u_0$, and iii) all the MBSs. By carefully handling these distributions, $\Pr\left[ \tau_{\textsf{s}_1}>\tau\right]$ in (\ref{psi1Def}) can be calculated.}
%{In (\ref{psi1Def}), $\Pr\left[ \tau_{\textsf{m}}>\tau\right]$} is the same as the coverage probability in the traditional connection-based HetNets and can be calculated using the method in \cite{Cache-enabledsmallcellnetworksmodelingandtradeoffs}. In calculating {$\Pr\left[ \tau_{\textsf{s}_1}>\tau\right]$}, we firstly need the joint probability density function (p.d.f.) of the distances of the $K$ SBSs in $\mathcal{C}_{1,n}$, which can be obtained by applying the method in \cite{10} and noting that $\Phi_{\textsf{s},n}$ is a homogeneous PPP with density $\lambda_{\textsf{s}} T_n$. Then, based on this joint p.d.f., {$\Pr\left[ \tau_{\textsf{s}_1}>\tau\right]$} can be calculated by following similar steps as in the derivation of the coverage probability in \cite{3}.
Thus, we have the following~theorem.
\newtheorem{Theorem}{Theorem}
\begin{Theorem}[STP under Scheme 1]\label{theorem1}
{The STP ${\psi_{\mathrm{sch}_1}(\mathbf{T})}$ of $u_0$ is given~by}
\begin{equation}\label{eqpsi1T}
{\psi_{\mathrm{sch}_1}(\mathbf{T})} = \sum\nolimits_{n \in \mathcal{N}} {a_n}({{\psi_{\textsf{m}}}\mathbbm{1}\left[T_n=0\right]  + {\psi_{\textsf{s}_1}(T_n)}\mathbbm{1}\left[T_n>0\right]}),
\end{equation}where $\psi_{\textsf{m}}$ and $\psi_{\textsf{s}_1}(T_n)$ {are given by
{\begingroup\makeatletter\def\f@size{9.5}\check@mathfonts
\def\maketag@@@#1{\hbox{\m@th\normalsize\normalfont#1}}\setlength{\arraycolsep}{0.0em}
\begin{align}
\psi_{\textsf{m}} &= \int\nolimits_0^\infty{\exp\left(-B_{\textsf{m},\textsf{s}}( \alpha_{\textsf{m}},\alpha_{\textsf{s}},1,\theta_{\textsf{m}},u)\right)}\mathrm{d}u,\label{eqPsim}\\
\psi _{\textsf{s}_1}\left( T_n\right) &= \begin{cases}\displaystyle
\displaystyle{\int\nolimits_0^\infty} \exp\left(-B_{\textsf{s},\textsf{m}}(\alpha_{\textsf{s}}, \alpha_{\textsf{m}},T_n,\theta_{\textsf{s}},u)\right)\mathrm{d}u,&K=1, \\
\displaystyle \int\nolimits_0^1\cdots\int\nolimits_0^1\int\nolimits_0^{\infty} \exp\left(-B_{\textsf{s},\textsf{m}}\left(\alpha_{\textsf{s}}, \alpha_{\textsf{m}},T_n,\frac{\theta_{\textsf{s}}}{1+\sum\nolimits_{k=1}^{K-1}t_k^{-\frac{\alpha_{\textsf{s}}}{2}}},u\right)\right)\frac{u^{K-1}}{(K-1)!} \mathrm{d}t_1\cdots\mathrm{d}t_{K-1}\mathrm{d}u, &K\ge2.
                                 \end{cases}\label{eqPsiS1}
\end{align}
\setlength{\arraycolsep}{5pt}\endgroup}Here, $\theta_{\textsf{m}}  \triangleq {2^{\tau /W_{\textsf{m}}}} - 1$, $\theta_{\textsf{s}}  \triangleq {2^{\tau /W_{\textsf{s}}}} - 1$ and
{\begingroup\makeatletter\def\f@size{10}\check@mathfonts
\def\maketag@@@#1{\hbox{\m@th\normalsize\normalfont#1}}\setlength{\arraycolsep}{0.0em}
\begin{eqnarray}\label{eqscheme1Bxy}
&&B_{x,y}(\alpha_{x},\alpha_{y},T,\theta,u)\triangleq \frac{2\pi ^{2 - {\alpha_{x}}/{\alpha_{y}}}}{\alpha_{y}}\csc \left(\frac{2\pi} {\alpha_{y}}\right)\frac{\lambda_{y}}{\lambda_{x}^{ {\alpha_{x}}/{\alpha_{y}}}}{\left( {\frac{\theta P_{y}}{P_x} } \right)^{2/\alpha_{y}}}\left(\frac{u}{T}\right)^{{\alpha_{x}}/{\alpha_{y}}}\nonumber\\
&&\quad\quad\quad\quad{+}\: u\left(\left(\frac{1}{T}-1\right) \frac{2\pi}{\alpha_{x}}\csc\left(\frac{2\pi}{\alpha_{x}}\right)
\theta^{{2}/{\alpha_{x}}}+{_2F_1}\left(-\frac{2}{\alpha_{x}},1;1-\frac{2}{\alpha_{x}};-\theta\right)\right),
\end{eqnarray}\setlength{\arraycolsep}{5pt}\endgroup}where $(x,y)=(\textsf{s},\textsf{m})$ or $(\textsf{m},\textsf{s})$ and $_2F_1(a,b;c;d)$ denotes the Gaussian hypergeometric function~\cite{Tableofintegralsseriesandproducts}.}\footnote{{Note that, when $\alpha=4$, we have ${_2F_1}(-\frac{2}{\alpha},1;1-\frac{2}{\alpha};-\theta)=\sqrt{\theta}\arctan(\sqrt{\theta})+1$.} }
\end{Theorem}

\indent\indent \emph{Proof}: See Appendix A.
%\begin{Remark}[Properties from Theorem \ref{theorem1}]\label{remark2}
%From Theorem \ref{theorem1}, we can easily see that the following properties hold.
%(i) As $\lambda_{\textsf{m}}$ increases, $\psi_{\textsf{m}}$ increases and $\psi_{\textsf{s}_1}(T_n)$ decreases;
%(ii) As $\lambda_{\textsf{s}}$ increases, $\psi_{\textsf{m}}$ decreases and $\psi_{\textsf{s}_1}(T_n)$ increases;
%(iii) As $T_n$ increases, $\psi_{\textsf{s}_1}(T_n)$ increases.
%\end{Remark}

In Theorem \ref{theorem1}, $\psi_{\textsf{m}}$ represents the STP of each file when $u_0$ is served by its serving MBS and $\psi_{\textsf{s}_1}(T_n)$ represents the STP of file $n$ when $u_0$ is jointly served by its serving SBSs in $\mathcal{C}_{1,n}$. Based on Theorem~\ref{theorem1}, we have the following remark.
\begin{Remark}[Properties of Theorem \ref{theorem1}]\label{remarkInterpretationTh1}
 From Theorem \ref{theorem1}, a few observations are in order.{
\begin{enumerate}
    \item $\psi_{\textsf{m}}$ increases with {\scriptsize$\frac{\lambda_{\textsf{m}}}{\lambda_{\textsf{s}}^{\alpha_{\textsf{s}}/\alpha_{\textsf{m}}}}$}. That is, the STP of a file transmitted by the nearest MBS is higher when the MBS density is larger or the SBS density is smaller.
    \item $\psi_{\textsf{s}_1}(T_n)$ increases with {\scriptsize$\frac{\lambda_{\textsf{s}}}{\lambda_{\textsf{m}}^{\alpha_{\textsf{m}}/\alpha_{\textsf{s}}}}$}. That is, the STP of file $n$ transmitted by the serving SBSs in $\mathcal{C}_{1,n}$ is higher when the SBS density is larger or the MBS density is smaller.
    \item $\psi_{\textsf{s}_1}(T_n)$ is an increasing function of $T_n$. That is, the STP of file $n$ transmitted by the serving SBSs in $\mathcal{C}_{1,n}$ is higher when the probability of storing file $n$ at an SBS is larger.
\end{enumerate}}
\end{Remark}

{Next, to obtain some simpler expressions for ${\psi_{\mathrm{sch}_1}(\mathbf{T})}$ in Theorem~\ref{theorem1}, we consider the symmetric case where $\alpha_{\textsf{s}}=\alpha_{\textsf{m}}=\alpha$. We have the following corollary.}
\begin{Corollary}[STP under Scheme 1 for $\alpha_{\textsf{s}}=\alpha_{\textsf{m}}=\alpha$]\label{corollaryScheme1K1}
When $\alpha_{\textsf{s}}=\alpha_{\textsf{m}}=\alpha$, the STP ${\psi_{\mathrm{sch}_1}(\mathbf{T})}$ is given by (\ref{psi1Def}) with $\psi_{\textsf{m}}$ and $\psi_{\textsf{s}_1}(T_n)$ given by
{\begingroup\makeatletter\def\f@size{10}\check@mathfonts
\def\maketag@@@#1{\hbox{\m@th\normalsize\normalfont#1}}\setlength{\arraycolsep}{0.0em}
\begin{eqnarray}\label{eqSec22close}
{\psi_{\textsf{m}}} &=& \frac{1}{B_{\textsf{m},\textsf{s}}(\alpha,\alpha,1,\theta_{\textsf{m}},1)},\\
{\psi_{\textsf{s}_1}(T_n)} &=&\displaystyle{\begin{cases}\displaystyle
                       \frac{1}{B_{\textsf{s},\textsf{m}}(\alpha,\alpha,T_n,\theta_{\textsf{s}},1)}, & K=1, \\
                       \displaystyle{\int\nolimits_0^1\cdots\int\nolimits_0^1}\frac{\mathrm{d}t_1\cdots\mathrm{d}t_{K-1}}{\left(B_{\textsf{s},\textsf{m}}\left( \alpha,\alpha,T_n, \frac{\theta_{\textsf{s}}}{1+\sum\nolimits_{k=1}^{K-1}t_k^{-\frac{\alpha}{2}}},1\right)\right)^{K}}, & K\ge2,
                     \end{cases}}
\end{eqnarray}\setlength{\arraycolsep}{5pt}\endgroup}where $B_{x,y}(\alpha_{x},\alpha_{y},T,\theta,u)$ is given by~(\ref{eqscheme1Bxy}).
\end{Corollary}

\indent\indent \emph{Proof}: Corollary \ref{corollaryScheme1K1} can be easily proved by using the equality $\int_{0}^{\infty}e^{-ax}x^{n-1}\mathrm{d}x=a^{-n}(n-1)!$. We omit the details due to page limitation.

{In {Corollary} \ref{corollaryScheme1K1}, we obtain a closed-form expression of $\psi_{\mathrm{sch}_1}(\mathbf{T})$ when $K=1$ and $\alpha_{\textsf{s}}=\alpha_{\textsf{m}}=\alpha$. In this case, ${\psi_{\textsf{s}_1}(T_n)}$ is concave and thus $\psi_{\mathrm{sch}_1}(\mathbf{T})$ is concave. Later, we shall see that the concavity greatly facilitates the optimization of $\psi_{\mathrm{sch}_1}(\mathbf{T})$ when $K=1$ and $\alpha_{\textsf{s}}=\alpha_{\textsf{m}}=\alpha$. Furthermore, when $K\ge2$ and $\alpha_{\textsf{s}}=\alpha_{\textsf{m}}=\alpha$, we obtain a simpler expression of  $\psi_{\mathrm{sch}_1}(\mathbf{T})$ than that of Theorem~1, which can be used to facilitate the numerical evaluation of $\psi_{\mathrm{sch}_1}(\mathbf{T})$.}

\begin{figure}[!t]
    \centering
    \subfloat{\includegraphics[width=2.5in]{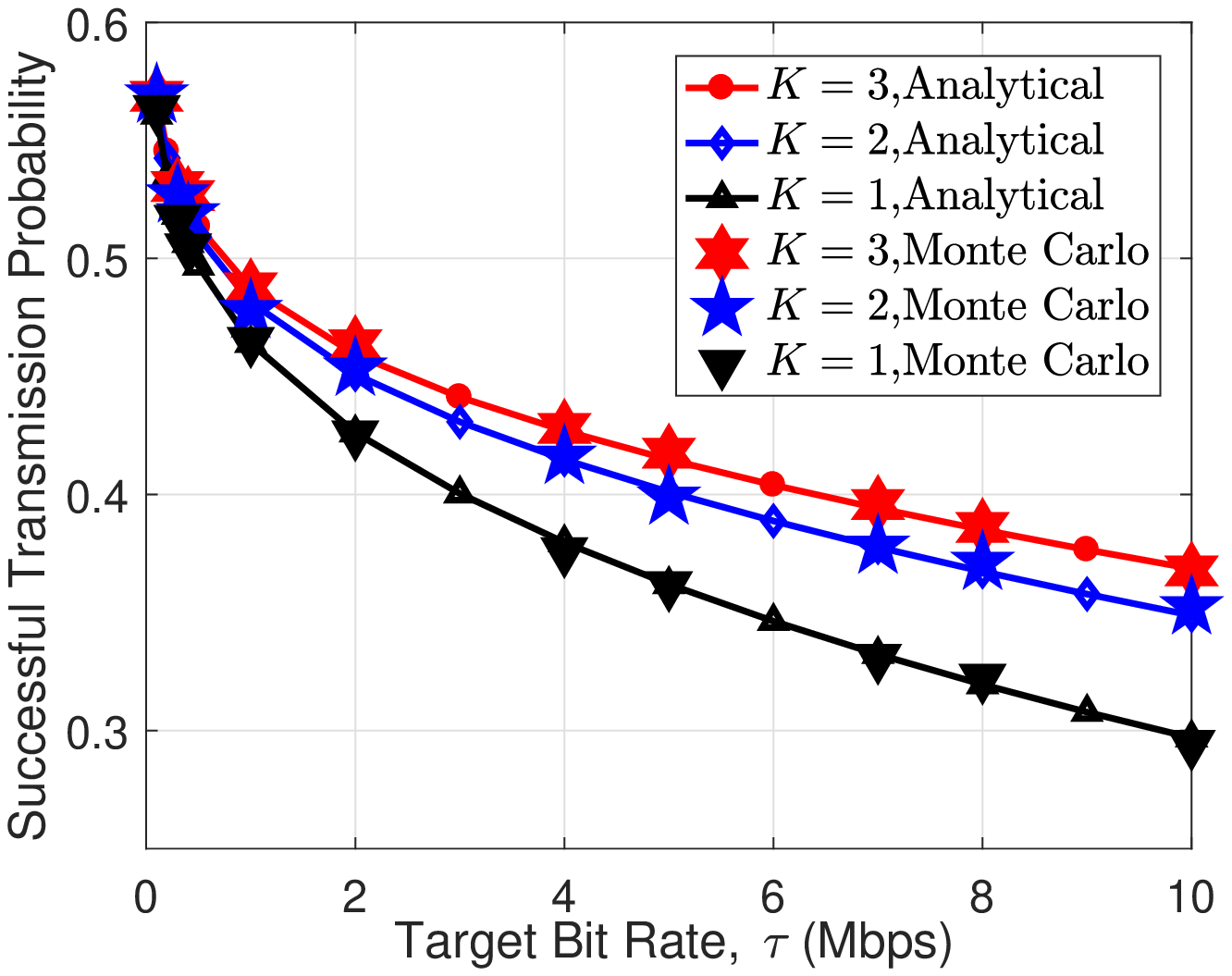}}
      \caption{STP ${\psi_{\mathrm{sch}_1}(\mathbf{T})}$ versus $\tau$. $M=2$, $N=10$, $T_1=0.9,T_2=0.8,T_3=0.3$, $T_n=0$ for $n=4,5,\cdots,N$, $\lambda_{\textsf{m}}=\frac{1}{ 500^2\pi}$ $\mathrm{m}^{-2}$, $\lambda_{\textsf{s}}=\frac{1}{ 50^2\pi}$ $\mathrm{m}^{-2}$, $P_{\textsf{m}}=43$ dBm, $P_{\textsf{s}}=23$ dBm, $\alpha_{\textsf{s}}=\alpha_{\textsf{m}}=4$, $W_{\textsf{m}}=0.2$ MHz, $W_{\textsf{s}}=20$ MHz, and $a_n=\frac{n^{-\gamma}}{\sum_{n \in\mathcal{N}}n^{-\gamma }}$ with Zipf exponent $\gamma=0.8$. In the Monte Carlo simulations, we choose a large spatial window, which is a square of $10^4 \times 10^4 $ $\mathrm{m}^2$, and the final simulation results are obtained by averaging over $10^4$ independent realizations.}\label{figpsi1}
\end{figure}

Fig. \ref{figpsi1}(a) plots the STP $\psi_{\mathrm{sch}_1}(\mathbf{T})$ versus the target bit rate $\tau$ and verifies Theorem~\ref{theorem1}. In addition, as expected, we see that the STP $\psi_{\mathrm{sch}_1}(\mathbf{T})$ decreases with the target bit rate $\tau$ and increases with the number of cooperative SBSs $K$. Besides, the marginal STP gain of including one more SBS into joint transmission decreases with $K$.
\subsection{Optimization of Successful Transmission Probability}\label{subsecoptimizationofpsisch1}
The caching distribution $\mathbf{T}$ significantly affects the STP under Scheme 1. We would like to maximize $\psi_{\mathrm{sch}_1}(\mathbf{T})$ in (\ref{eqpsi1T}) by optimizing $\mathbf{T}$. Note that, when studying Scheme 1, we focus on the region in which $\psi_{\textsf{s}_1}({1})>\psi_{\textsf{m}}$. In this region, $u_0$ prefers receiving files from SBSs, as SBSs can offer a higher STP than the nearest MBS. Specifically, we have the following~problem.
\newtheorem{Problem}{Problem}
\begin{Problem}[Optimization of STP under Scheme 1]\label{problem1}
\setlength{\arraycolsep}{0.0em}
\begin{eqnarray}
\psi_{\mathrm{sch}_1}^* &\triangleq& \mathop {\max }\limits_{ {\mathbf{T}}} \;\;\psi_{\mathrm{sch}_1}(\mathbf{T})\label{p1}\\
&&s.t.\;\;(\ref{Tncons1}), (\ref{Tncons2}).\nonumber
\end{eqnarray}\setlength{\arraycolsep}{5pt}Here, $\mathbf{T}^*$ denotes the optimal solution and $\psi_{\mathrm{sch}_1}^*=\psi_{\mathrm{sch}_1}(\mathbf{T}^*)$ denotes the optimal~value.
\end{Problem}

As the structure of $\psi_{\mathrm{sch}_1}(\mathbf{T})$ in {Problem} \ref{problem1} is very complex. To obtain design insights, we first analyze the optimality properties of {Problem} \ref{problem1}.
\begin{Lemma}[Optimality Properties of Problem \ref{problem1}]\label{lemmaOptimalityPropertiesp11}
There exists $N_{\textsf{s}}^*\in\Big\{M,M+1,\cdots,\min\Bigl\{\left\lceil\frac{M}{T_{\mathrm{th}}}\right\rceil-1,N\Bigr\}\Big\}$ such that the optimal solution $\mathbf{T}^*$ to {Problem} \ref{problem1} satisfies $1\ge T_1^*\ge T_2^*\ge\cdots\ge T_{N_{\textsf{s}}^*}^*> T_{\mathrm{th}}$ and $T_{N_{\textsf{s}}^*+1}^*= T_{N_{\textsf{s}}^*+2}^*=\cdots= T_{N}^*=0$, where $T_{\mathrm{th}}\in (0,1)$ is the root to {$\psi_{\textsf{s}_1}(x)=\psi_{\textsf{m}}$.}\footnote{$T_{\mathrm{th}}$ can be calculated by the bisection method due to the monotonicity of $\psi_{\textsf{s}_1}(x)$ w.r.t. $x$.}
\end{Lemma}

\indent\indent {\emph{Proof}}: See Appendix B.
\begin{Remark}[Interpretation of Lemma \ref{lemmaOptimalityPropertiesp11}]\label{remarkInterpretationlemmaOptimalityPropertiesp11}
From {Lemma} \ref{lemmaOptimalityPropertiesp11}, a few observations are in  order.
\begin{enumerate}
\item A file of higher popularity {should be stored} at the SBS tier with a higher probability (i.e., stored at more SBSs), and some files of low popularity may not be stored. In addition, the $N_{\textsf{s}}^*$ most popular files are stored at the SBS tier and their caching probabilities are no smaller than~$T_{\mathrm{th}}$.{
%\item$\min\left\{\left\lceil\frac{M}{T_{\mathrm{th}}}\right\rceil-1,N\right\}$ is nondecreasing with the decrease of $T_{\mathrm{th}}$, implying that more files should be stored at the SBS tier with decreasing $T_{\mathrm{th}}$.
\item From (\ref{eqPsim}), we see that $\psi_{\textsf{m}}$ is independent of $K$. From (\ref{eqgammas1n}) and (\ref{psi1Def}), we know that $\psi_{\textsf{s}_1}(T_n)$ increases with $K$. Hence, as $K$ increasing, the root of equation $\psi_{\textsf{s}_1}(x)=\psi_{\textsf{m}}$, i.e., $T_{\mathrm{th}}$, decreases with $K$, implying that $\min\left\{\left\lceil\frac{M}{T_{\mathrm{th}}}\right\rceil-1,N\right\}$ is nondecreasing with $K$. That is, as more SBSs jointly serving $u_0$, more files can be stored at the SBS tier.}
%\item From Theorem \ref{theorem1}, we know that $\psi_{\textsf{s}_1}(T_n)$ is an increasing function of $T_n$, and $\psi_{\textsf{m}}$ is independent of $K$. In addition, from (\ref{eqgammas1n}) and (\ref{psi1Def}) , we know that $\psi_{\textsf{s}_1}(T_n)$ increases with $K$. Hence, the root of equation $\psi_{\textsf{s}_1}(T_n)=\psi_{\textsf{m}}$, i.e., $T_{\mathrm{th}}$, decreases with $K$, implying that $\min\left\{\left\lceil\frac{M}{T_{\mathrm{th}}}\right\rceil-1,N\right\}$ is nondecreasing with $K$. In other words, as $K$ increases, more files {should be stored at} the SBS tier.
\item When $N_{\textsf{s}}^*=M$, we have $T_n^*=1$ for $n=1,\cdots,M$, and $T_n^*=0$ for $n=M+1,M+2,\cdots,N$, i.e., the optimal caching reduces to MPC \cite{Cache-enabledsmallcellnetworksmodelingandtradeoffs}.
\end{enumerate}
\end{Remark}

In general, it is difficult to show the convexity of the objective function $\psi_{\mathrm{sch}_1}(\mathbf{T})$. However, the constraint set is obviously convex. In addition, due to the indicator function $\mathbbm{1}[\bullet]$ in $\psi_{\mathrm{sch}_1}(\mathbf{T})$, the objective function $\psi_{\mathrm{sch}_1}(\mathbf{T})$ is not differentiable w.r.t. $\mathbf{T}$, which means that we cannot directly apply the standard gradient projection method in \cite{CuiCacheWirelessSignleTier} to obtain a local optimal solution of {Problem} \ref{problem1} numerically. In the following, we construct an equivalent problem of {Problem} \ref{problem1} by making use of the optimality properties in {Lemma} \ref{lemmaOptimalityPropertiesp11}.

From {Lemma}~\ref{lemmaOptimalityPropertiesp11}, we know that the $N_{\textsf{s}}^*$ most popular files are stored in the SBS tier. Thus, to solve {Problem} \ref{problem1}, we can introduce an auxiliary variable $N_{\textsf{s}}\in\Big\{M,M+1,\cdots,\min\Bigl\{\left\lceil\frac{M}{T_{\mathrm{th}}}\right\rceil-1,N\Bigr\}\Big\}$ and rewrite the STP $\psi_{\mathrm{sch}_1}(\mathbf{T})$ in (\ref{psi1Def}) as
\setlength{\arraycolsep}{0.0em}
\begin{eqnarray}\label{eqrewrittenPsiTNs}
    \psi_{\mathrm{sch}_1}(\mathbf{T},N_{\textsf{s}}) &=& \mathcal{P}_{\textsf{s}}(\mathbf{T},N_{\textsf{s}}) + \mathcal{P}_{\textsf{m}}(N_{\textsf{s}}),
\end{eqnarray}\setlength{\arraycolsep}{5pt}where $\mathcal{P}_{\textsf{s}}(\mathbf{T},N_{\textsf{s}})\triangleq\sum_{n=1}^{N_{\textsf{s}}}a_n{\psi_{\textsf{s}_1}(T_n)}$, $\mathcal{P}_{\textsf{m}}(N_{\textsf{s}})\triangleq\sum_{n=N_{\textsf{s}}+1}^{N}a_n{\psi_{\textsf{m}}}$, $T_n\ge T_{\mathrm{th}}$ for $n=1,\cdots,N_{\textsf{s}}$, and $T_n=0$ for $n=N_{\textsf{s}}+1,N_{\textsf{s}}+2,\cdots,N$. Note that $\psi_{\mathrm{sch}_1}(\mathbf{T},N_{\textsf{s}})$ is differentiable w.r.t. $\mathbf{T}$, for any given $N_{\textsf{s}}$. Thus, we have an equivalent problem of {Problem} \ref{problem1} as follows.
\begin{Problem}[Equivalent Problem of Problem \ref{problem1}]\label{eqproblem1}
\setlength{\arraycolsep}{0.0em}
\begin{eqnarray}
\psi_{\mathrm{sch}_1}^* &\triangleq& \mathop {\max }\limits_{ {\mathbf{T},N_{\textsf{s}}}} \;\;\psi_{\mathrm{sch}_1}(\mathbf{T},N_{\textsf{s}})\label{sp11}\nonumber\\
s.t.\;\;&& N_{\textsf{s}}\in\left\{M,M+1,\cdots,\min\left\{\left\lceil\frac{M}{T_{\mathrm{th}}}\right\rceil-1,N\right\}\right\},\label{eqPorbCond1forNs}\\
&&T_{\mathrm{th}}\le T_n\le1,\;n=1,\cdots,N_{\textsf{s}},\label{newCondforTnSBSs}\\
&&T_n=0,\;n=N_{\textsf{s}}+1,N_{\textsf{s}}+2,\cdots,N,\label{newCondforTnMBS}\\
&&\sum_{n=1}^{N_{\textsf{s}}}T_n=M.\label{eqPorbCond3}
\end{eqnarray}
\setlength{\arraycolsep}{5pt}Here, $\mathbf{T}^*$ and $N_{\textsf{s}}^*$ denote the optimal solution and $\psi_{\mathrm{sch}_1}^*=\psi_{\mathrm{sch}_1}(\mathbf{T}^*,N_{\textsf{s}}^*)$ denotes the optimal value.
\end{Problem}

Note that $\mathbf{T}^*$ and $\psi_{\mathrm{sch}_1}^*$ given by Problem \ref{eqproblem1} are the same as those given by Problem \ref{problem1}. Therefore, instead of solving {Problem} \ref{problem1}, we can solve {Problem} \ref{eqproblem1}. {Problem} \ref{eqproblem1} is a mixed discrete-continuous optimization problem with two main challenges. One is the choice of the number of different files stored at the SBS tier, i.e., $N_{\textsf{s}}$ (discrete variables), and the other is the choice of the caching distribution $\mathbf{T}$ (continuous variables) of the random caching scheme for the SBS tier. We thus propose an equivalent alternative formulation of {Problem} \ref{eqproblem1} which naturally subdivides {Problem} \ref{eqproblem1} according to these two aspects.

\begin{Problem}[Equivalent Problem of Problem \ref{eqproblem1}]\label{eqproblem2}
\setlength{\arraycolsep}{0.0em}
\begin{eqnarray}
\psi_{\mathrm{sch}_1}^* &\triangleq \mathop {\max }\limits_{{N_{\textsf{s}}}}\;\;\mathcal{P}_{\textsf{s}}^*(N_{\textsf{s}}) + \mathcal{P}_{\textsf{m}}(N_{\textsf{s}})\\
&s.t.\;\; (\ref{eqPorbCond1forNs}),\nonumber
\end{eqnarray}
\setlength{\arraycolsep}{5pt} where
\begin{eqnarray}\label{sp2}
\mathcal{P}_{\textsf{s}}^*(N_{\textsf{s}}) &\triangleq \mathop {\max }\limits_{{\mathbf{T}}}\;\;\mathcal{P}_{\textsf{s}}(\mathbf{T},N_{\textsf{s}}) \\
&s.t.\;\;(\ref{newCondforTnSBSs}),(\ref{newCondforTnMBS}),(\ref{eqPorbCond3}).\nonumber
\end{eqnarray}\setlength{\arraycolsep}{5pt}Here, $\mathbf{T}^*(N_{\textsf{s}})$ denotes the optimal solution to the optimization in (\ref{sp2}) for given $N_{\textsf{s}}$, $\mathcal{P}_{\textsf{s}}^*(N_{\textsf{s}})=\mathcal{P}_{\textsf{s}}^*(\mathbf{T}^*(N_{\textsf{s}}),N_{\textsf{s}})$ denotes the optimal value of the optimization in (\ref{sp2}) for given $N_{\textsf{s}}$, $N_{\textsf{s}}^*$ denotes the optimal solution to the optimization in (\ref{sp11}), and $\psi_{\mathrm{sch}_1}^*=\mathcal{P}_{\textsf{s}}^*(N_{\textsf{s}}^*)+\mathcal{P}_{\textsf{m}}(N_{\textsf{s}}^*)$ denotes the optimal value of the optimization in (\ref{sp11}). Note that $\mathbf{T}^*(N_{\textsf{s}}^*)=\mathbf{T}^*$, where $\mathbf{T}^*$ is given by Problem~\ref{eqproblem1}.
\end{Problem}

For given $N_{\textsf{s}}$, the problem in (\ref{sp2}) is a continuous optimization of a differentiable function $\mathcal{P}_{\textsf{s}}(\mathbf{T},N_{\textsf{s}})$ over a convex set. In general, it is difficult to show the convexity of $\psi_{\textsf{s}_1}(T_n)$ in (\ref{eqPsiS1}) and hence the convexity of $\mathcal{P}_{\textsf{s}}(\mathbf{T},N_{\textsf{s}})$. Since $\mathcal{P}_{\textsf{s}}(\mathbf{T},N_{\textsf{s}})$ is differentiable, we can apply the standard gradient projection method, e.g., Algorithm 1 in \cite{CuiCacheWirelessSignleTier}, to obtain a local optimal solution to the optimization in (\ref{sp2}).

%Due to the $K$-fold integral in $\psi_{\textsf{s}_1}(T_n)$, it is of huge complexity to calculate the gradient of $\mathcal{P}_{\textsf{s}}(\mathbf{T},N_{\textsf{s}})$ when $K$ is large. From {Lemma} \ref{lemma1upperboundforScheme1}, we know that when $K\ge2$ and $\alpha_{\textsf{s}}=\alpha_{\textsf{m}}=\alpha$, there exists a simpler approximate expression $\widetilde{\psi}_{\textsf{s}_1}(T_n)$ with a single-fold integral only. Thus, we can consider the optimization of $\widetilde{\mathcal{P}}_{\textsf{s}}(\mathbf{T},N_{\textsf{s}}) \triangleq\sum_{n=1}^{N_{\textsf{s}}}a_n{\widetilde{\psi}_{\textsf{s}_1}(T_n)}$, instead of $\mathcal{P}_{\textsf{s}}(\mathbf{T},N_{\textsf{s}})$ for $K\ge2$ and $\alpha_{\textsf{s}}=\alpha_{\textsf{m}}=\alpha$ as an approximation, for the purpose of reducing computation complexity in using the standard gradient projection method. Note that the above-mentioned properties for ${\psi}_{\textsf{s}_1}(T_n)$ also hold for $\widetilde{\psi}_{\textsf{s}_1}(T_n)$. Thus, the optimization of $\widetilde{\mathcal{P}}_{\textsf{s}}(\mathbf{T},N_{\textsf{s}})$ can be done in a similar~manner.
From {Corollary} \ref{corollaryScheme1K1}, we know that when $K=1$ and $\alpha_{\textsf{s}}=\alpha_{\textsf{m}}=\alpha$, $\psi_{\textsf{s}_1}(T_n)$ is concave, implying that $\mathcal{P}_{\textsf{s}}(\mathbf{T},N_{\textsf{s}})$ is concave, and Slater¡¯s condition is satisfied, implying that strong duality holds. In this case, we can obtain a closed-form optimal solution to the convex optimization in (\ref{sp2}) using KKT conditions.
\begin{Lemma}[Optimal Solution to Problem (\ref{sp2}) for $K=1$ and $\alpha_{\textsf{s}}=\alpha_{\textsf{m}}=\alpha$]\label{lemmasp2forK1closedfrom}
When $K=1$ and $\alpha_{\textsf{s}}=\alpha_{\textsf{m}}=\alpha$, for any given $N_{\textsf{s}}\le\min\left\{\left\lceil\frac{M}{T_{\mathrm{th}}}\right\rceil-1,N\right\}$, the optimal solution to the optimization in (\ref{sp2}) is given by
\setlength{\arraycolsep}{0.0em}
\begin{eqnarray*}
T_n^*(N_{\textsf{s}})=
\begin{cases}
\min \left\{ {\max \left\{ {\frac{1}{{{c_2}}}\left( {\sqrt {\frac{{{a_n}{c_1}}}{{{\nu ^*}}}}  - {c_1}} \right),{T_{\mathrm{th}}}} \right\},1} \right\},n=1,\cdots,N_{\textsf{s}}\\
0,\;n=N_{\textsf{s}}+1,N_{\textsf{s}}+2,\cdots,N
\end{cases}
\end{eqnarray*}
\setlength{\arraycolsep}{5pt}where $c_1\triangleq{2\pi}{\alpha^{-1}}\csc\left({2\pi}{\alpha^{-1}}\right)\theta_{\textsf{s}}^{2\alpha^{-1}} (1+\lambda_{\textsf{m}}\lambda_{\textsf{s}}^{-1}(P_{\textsf{m}}P_{\textsf{s}}^{-1})^{2\alpha^{-1}})$, $c_2\triangleq-{2\pi}{\alpha^{-1}}\csc\left({2\pi}{\alpha^{-1}}\right)\theta_{\textsf{s}}^{2\alpha^{-1}}
+{_2F_1}(-2\alpha^{-1},1;1-2\alpha^{-1};-\theta_{\textsf{s}})$, $c_3\triangleq{2\pi}{\alpha^{-1}}\csc\left({2\pi}{\alpha^{-1}}\right)\theta_{\textsf{m}}^{2\alpha^{-1}} (1+\lambda_{\textsf{s}}\lambda_{\textsf{m}}^{-1}(P_{\textsf{s}}P_{\textsf{m}}^{-1})^{2\alpha^{-1}})$, $T_{\mathrm{th}}=\frac{c_{1}}{c_{3}}$, and $\nu^*$ satisfies
\setlength{\arraycolsep}{0.0em}
\begin{eqnarray*}
&&\sum\limits_{n=1}^{N_{\textsf{s}}}\min \left\{ {\max \left\{ {\frac{1}{{{c_2}}}\left( {\sqrt {\frac{{{a_n}{c_1}}}{{{\nu ^*}}}}  - {c_1}} \right),{T_{\mathrm{th}}}} \right\},1} \right\}=M.\label{eqnustar}
\end{eqnarray*}
\setlength{\arraycolsep}{5pt}
\end{Lemma}

{\indent \indent \emph{Proof}: Lemma \ref{lemmasp2forK1closedfrom} can be proved in a similar way to Lemma 6 in \cite{AnalysisandOptimizationofCachingandMulticastinginLargeScaleCacheEnabledHeterogeneousWirelessNetworks}. We omit the details due to page limitation.}

\begin{algorithm}[!t]
\caption{\small Optimal Solution to the {Problem} \ref{eqproblem2}}
\begin{algorithmic}[1]\label{algorithm1}\small
\STATE $\psi_{\mathrm{sch}_1}^*\leftarrow 0$.{
\STATE Calculate $T_{\mathrm{th}}$ using Lemma~\ref{lemmasp2forK1closedfrom} (when $K=1$ and $\alpha_{\textsf{s}}=\alpha_{\textsf{m}}=\alpha$) or the bisection method (when $K\ge2$ or $\alpha_{\textsf{s}}\neq\alpha_{\textsf{m}}$).}
\FOR {$N_{\textsf{s}}^*=M$ to $\min\left\{\left\lceil\frac{M}{T_{\mathrm{th}}}\right\rceil-1,N\right\}$}
\STATE Obtain $\mathbf{T}^*(N_{\textsf{s}}^*)$ and $\mathcal{P}_{\textsf{s}}^*(N_{\textsf{s}}^*)$ by solving the optimization in (\ref{sp2}) using {Lemma} \ref{lemmasp2forK1closedfrom} (when $K=1$ and $\alpha_{\textsf{s}}=\alpha_{\textsf{m}}=\alpha$) or the gradient projection method (when $K\ge2$ or $\alpha_{\textsf{s}}\neq\alpha_{\textsf{m}}$).
\IF {$\psi_{\mathrm{sch}_1}^*<\mathcal{P}_{\textsf{s}}^*(N_{\textsf{s}}^*)+\mathcal{P}_{\textsf{m}}(N_{\textsf{s}}^*)$}
\STATE $\psi_{\mathrm{sch}_1}^*\leftarrow\mathcal{P}_{\textsf{s}}^*(N_{\textsf{s}}^*)+\mathcal{P}_{\textsf{m}}(N_{\textsf{s}}^*)$ and $\mathbf{T}^*\leftarrow \mathbf{T}^*(N_{\textsf{s}}^*)$
\ENDIF
\ENDFOR
\end{algorithmic}
\end{algorithm}

Given solutions obtained using the standard gradient projection method or Lemma \ref{lemmasp2forK1closedfrom}, the optimization in (\ref{sp11}) is a discrete optimization over the set $\left\{M,M+1,\cdots,\min\left\{\left\lceil\frac{M}{T_{\mathrm{th}}}\right\rceil-1,N\right\}\right\}$ of cardinality $\min\left\{\left\lceil\frac{M}{T_{\mathrm{th}}}\right\rceil-1,N\right\}-M+1$. The discrete optimization problem in (\ref{sp11}) can be solved directly using exhaustive search of complexity $\mathcal{O}(N)$.

Finally, combining the above discrete part and continuous part, we can obtain a global (when $K=1$ and $\alpha_{\textsf{s}}=\alpha_{\textsf{m}}=\alpha$) or local (when $K\ge2$ or $\alpha_{\textsf{s}}\neq\alpha_{\textsf{m}}$) optimal solution to {Problem}~\ref{eqproblem2} as summarized in Algorithm \ref{algorithm1}.
\section{Analysis and Optimization of Performance under Scheme 2}
In this section, we first analyze the STP under Scheme 2 for a given caching distribution $\mathbf{T}$ of the random caching scheme. Then we maximize the STP by optimizing $\mathbf{T}$.
\subsection{Analysis of Successful Transmission Probability}
%{\color{black}Note that, conditioning on $C_{2,n}=0$, $\Pr \left[ \tau_{\textsf{m}}|C_{2,n}=0}\right] = \Pr \left[ \tau_{\textsf{m}} > \tau\right]=\psi_{\textsf{m}}$ which is already given by (\ref{eqPsim}). However, conditioning on $C_{2,n}\ge1$, it is challenging to compute $\Pr\left[ \tau_{\textsf{s}_2}>\tau|C_{2,n}=k\right]$, since the joint p.d.f. of the distances between the $k$ serving SBSs and $u_0$ is still undetermined.}
In this part, we analyze the STP $\psi_{\mathrm{sch}_2}(\mathbf{T})$ in (\ref{psi2Def}), using tools from stochastic geometry. {It is more challenging to calculate $\psi_{\mathrm{sch}_2}(\mathbf{T})$ than to calculate $\psi_{\mathrm{sch}_1}(\mathbf{T})$ under Scheme 1, as the number of the serving BSs of $u_0$, i.e., $C_{2,n}$, is a random variable. Specifically,} if $C_{2,n}=0$, $u_0$ is served by its nearest MBS, or otherwise by the $C_{2,n}$ SBSs in $\mathcal{C}_{2,n}$. Thus, to calculate $\psi_{\mathrm{sch}_2}(\mathbf{T})$, we first need to calculate the probability mass function (p.m.f.) $\mathrm{Pr} \left[C_{2,n}=k\right]$, $k=0,1,\cdots,K$ of $C_{2,n}$. Under the random caching scheme, each SBS stores file $n$ with probability $T_n$, and we have $\mathcal{C}_{2,n}\subseteq\mathcal{C}_2$ and $|\mathcal{C}_2|=K$. Thus, $C_{2,n}$ follows the binomial distribution with parameter $K$ and $T_n$, i.e., $\mathrm{Pr}\left[C_{2,n}=k \right]= \binom{K}{k}T_n^k(1-T_n)^{K-k}$. %However, given $C_{2,n}=k\ge1$, it is still challenging to determine the joint p.d.f. of the distances between the $k$ serving SBSs and $u_0$. To calculate this p.d.f., we consider the following two cases: i) there exist $k$ SBSs out of the $K-1$ nearest SBSs storing file $n$ and the $K$-th nearest SBS does not store file $n$; ii) there exist $k-1$ SBSs out of the $K-1$ nearest SBSs storing file $n$ and the $K$-th nearest SBS stores file $n$. By carefully handling these two cases, the joint p.d.f. of the distances between the $k$ SBSs and $u_0$ can be obtained. Thus, we have the following theorem.
{Note that, we have $\Pr \left[ \tau_{\textsf{m}}>\tau,C_{2,n}=0\right] = \Pr \left[ \tau_{\textsf{m}} > \tau\right]\Pr[C_{2,n}=0]$ and $\Pr\left[ \tau_{\textsf{s}_2}>\tau,C_{2,n}=k\right]=\Pr[ \tau_{\textsf{s}_2}>\tau|C_{2,n}=k]\Pr[C_{2,n}=k]$. Since $\Pr[\tau_{\textsf{m}}>\tau]$ is already given by Theorem \ref{theorem1}, it remains to calculate $\Pr[ \tau_{\textsf{s}_2}>\tau|C_{2,n}=k]$, which depends on the joint p.d.f. of the distances between the $k$ serving SBSs and $u_0$. To calculate this joint p.d.f.,} we consider the following two cases: i) there exist $k$ SBSs out of the $K-1$ nearest SBSs storing file $n$ and the $K$-th nearest SBS does not store file $n$; ii) there exist $k-1$ SBSs out of the $K-1$ nearest SBSs storing file $n$ and the $K$-th nearest SBS stores file $n$. By carefully handling these two cases, the joint p.d.f. of the distances between the $k$ SBSs and $u_0$ can be obtained. {\color{black}Then, based on this joint p.d.f., {$\Pr\left[ \tau_{\textsf{s}_2}>\tau|C_{2,n}=k\right]$} can be calculated by following similar steps as in the derivation of $\psi _{\textsf{s}_1}\left( T_n\right)$ in (\ref{eqPsiS1})}. Thus, we have the following theorem.
\begin{Theorem}[STP under Scheme 2]\label{theorem2}
The STP $\psi_{\mathrm{sch}_2}(\mathbf{T})$ of $u_0$ is given by
\setlength{\arraycolsep}{0.0em}
\begin{equation}\label{eqnewpsi2T}
\psi_{\mathrm{sch}_2}(\mathbf{T}) = \sum\nolimits_{n \in {\cal N}} a_n\psi_{\textsf{ms}}(T_n),
\end{equation}\setlength{\arraycolsep}{5pt}where $\psi_{\textsf{ms}}(T_n)$ is given by
{\begingroup\makeatletter\def\f@size{10}\check@mathfonts
\def\maketag@@@#1{\hbox{\m@th\normalsize\normalfont#1}}\setlength{\arraycolsep}{0.0em}
\begin{eqnarray}\label{eqscheme2psims}
\psi_{\textsf{ms}}(T_n)\triangleq (1-T_n)^K\psi_{\textsf{m}} +\sum\nolimits_{k = 1}^K {\binom{K}{k}}{T_n^k{\left( {1 - {T_n}} \right)^{K - k}}}\psi_{\textsf{s}_2,k}.
\end{eqnarray}\setlength{\arraycolsep}{5pt}\endgroup}Here, $\psi_{\textsf{m}}$ is given by (\ref{eqPsim}) and $\psi_{\textsf{s}_2,k}\triangleq \left(1-\frac{k}{K}\right)q_{k,1}+\frac{k}{K}q_{k,2}$ with $q_{k,1}$ and $q_{k,2}$ given by{
{\begingroup\makeatletter\def\f@size{10}\check@mathfonts
\def\maketag@@@#1{\hbox{\m@th\normalsize\normalfont#1}}\setlength{\arraycolsep}{0.0em}
\begin{align}
% \nonumber to remove numbering (before each equation)
{q_{k,1}} &=
{
\begin{cases}\label{eqPsik1}
  \displaystyle\int\nolimits_0^1\cdots\int\nolimits_0^1\int\nolimits_0^{\infty}\exp\left(-B_{\textsf{s},\textsf{m}}\left(\alpha_{\textsf{s}},\alpha_{\textsf{m}},1,\frac{\theta_{\textsf{s}}}{\sum\nolimits_{i=1}^kt_i^{-\frac{\alpha_{\textsf{s}}}{2}}},u\right)\right)\frac{{u^{K - 1}}}{{(K - 1)!}}\mathrm{d}t_1\cdots\mathrm{d}t_k\mathrm{d}{u}, &  k=1,\cdots,K-1, \\
  0, &  k=K,
\end{cases}
 }\\
{q_{k,2}} &=  \begin{cases}\label{eqPsik2}
\displaystyle \int_0^\infty \exp \left(-B_{\textsf{s},\textsf{m}}\left( \alpha_{\textsf{s}},\alpha_{\textsf{m}},1,{\theta_{\textsf{s}}},u\right)\right)\frac{{u^{K - 1}}}{{(K - 1)!}}\mathrm{d}{u}, & k=1,\\
 \displaystyle \int\nolimits_0^1\cdots\int\nolimits_0^1\int\nolimits_0^{\infty}  \exp \left(-B_{\textsf{s},\textsf{m}}\left( \alpha_{\textsf{s}},\alpha_{\textsf{m}},1,\frac{\theta_{\textsf{s}}}{1+\sum\nolimits_{i=1}^kt_i^{-\frac{\alpha_{\textsf{s}}}{2}}},u\right)\right)\frac{{u^{K - 1}}}{{(K - 1)!}}\mathrm{d}t_1\cdots\mathrm{d}t_{k-1}\mathrm{d}{u}, & k=2,\cdots,K,
               \end{cases}
\end{align}\setlength{\arraycolsep}{5pt}\endgroup}where $B_{x,y}(\alpha_{x},\alpha_{y},T,\theta,u)$ is given by (\ref{eqscheme1Bxy}).}
\end{Theorem}

\indent\indent {\emph{Proof}}: See Appendix C.
%\begin{Remark}[Properties from Theorem 2]\label{propTh2Scheme2}
%From Theorem 2, the following properties hold. {\color{black}i) $\psi_{\textsf{s}_2,k}$ increases with $k$; ii) If $\psi_{\textsf{m}}<\psi_{\textsf{s}_2,1}$, $\psi_{\textsf{ms}}(T_n)$ is a non-decreasing function of $T_n$; {\color{blue}iii) If $K=1$, $\psi_{\textsf{ms}}(T_n)$ is linear and thus, $\psi_{\mathrm{sch}_2}(\mathbf{T})$ is linear; iv)} If $K\ge2$ and $\psi_{\textsf{s}_2,k+1}-\psi_{\textsf{s}_2,k}\le\psi_{\textsf{s}_2,k}-\psi_{\textsf{s}_2,k-1}$ for all $k=1,\cdots,K-1$ where $\psi_{\textsf{s}_2,0}\triangleq\psi_{\textsf{m}}$,} $\psi_{\textsf{ms}}(T_n)$ is concave \cite{DistributedCachingandSmallCellCooperationforFastContentDelivery} and thus, $\psi_{\mathrm{sch}_2}(\mathbf{T})$ is concave. Note that, the condition $\psi_{\textsf{s}_2,k+1}-\psi_{\textsf{s}_2,k}\le\psi_{\textsf{s}_2,k}-\psi_{\textsf{s}_2,k-1}$ implies that the marginal STP gain of including one more SBS into joint transmission is decreasing and is smaller than the marginal STP gain of using an SBS instead of the nearest~MBS.
%\end{Remark}

In Theorem \ref{theorem2}, $\psi_{\textsf{ms}}(T_n)$ represents the STP of file $n$ and $\psi_{\textsf{s}_2,k}$ represents the conditional STP of file $n$, given that $u_0$ is jointly served by the $C_{2,n}=k$ SBSs in $\mathcal{C}_{2,n}$. Based on Theorem 2, we have the following remark.
\begin{Remark}[Properties of Theorem 2]\label{propTh2Scheme2} From Theorem 2, a few observations are in order.{
\begin{enumerate}
  \item From (\ref{eqscheme2psims}), (\ref{eqPsik1}) and (\ref{eqPsik2}), we easily see that $\psi_{\textsf{s}_2,k}>\psi_{\textsf{s}_2,k-1}$ for all $k=2,\cdots,K$, which means that including one more SBS into joint transmission yields a higher STP.
  \item If $\psi_{\textsf{s}_2,1}>\psi_{\textsf{m}}$, $\psi_{\textsf{ms}}(T_n)$ is an increasing function of $T_n$. That is, a file of higher probability being stored at an SBS has a higher STP. Furthermore, if $\psi_{\textsf{s}_2,k+1}-\psi_{\textsf{s}_2,k}\le\psi_{\textsf{s}_2,k}-\psi_{\textsf{s}_2,k-1}$ for all $k=1,\cdots,K-1$, where $\psi_{\textsf{s}_2,0}\triangleq\psi_{\textsf{m}}$, $\psi_{\textsf{ms}}(T_n)$ is concave \cite{DistributedCachingandSmallCellCooperationforFastContentDelivery} and thus, $\psi_{\mathrm{sch}_2}(\mathbf{T})$ is concave. Later, we shall see that the concavity greatly facilitates the optimization of the STP $\psi_{\mathrm{sch}_2}(\mathbf{T})$. Note that, the condition $\psi_{\textsf{s}_2,k+1}-\psi_{\textsf{s}_2,k}\le\psi_{\textsf{s}_2,k}-\psi_{\textsf{s}_2,k-1}$ implies that the marginal STP gain of including one more SBS into joint transmission is decreasing and is smaller than the marginal STP gain of using an SBS instead of the nearest~MBS.
\item If $K=1$, $\psi_{\textsf{ms}}(T_n)$ is linear and thus, $\psi_{\mathrm{sch}_2}(\mathbf{T})$ is linear. Later, we shall see that the linearity greatly facilitates the optimization of the STP $\psi_{\mathrm{sch}_2}(\mathbf{T})$ when $K=1$.
\end{enumerate}}
\end{Remark}

{Next, to obtain some simpler expressions for ${\psi_{\mathrm{sch}_2}(\mathbf{T})}$ in Theorem~\ref{theorem2}, we consider the symmetric case where $\alpha_{\textsf{s}}=\alpha_{\textsf{m}}=\alpha$. We have the following corollary.}

\begin{Corollary}[STP under Scheme 2 for $\alpha_{\textsf{s}}=\alpha_{\textsf{m}}=\alpha$]\label{corollaryScheme2}
When $\alpha_{\textsf{s}}=\alpha_{\textsf{m}}=\alpha$, the STP ${\psi_{\mathrm{sch}_2}(\mathbf{T})}$ is given by (\ref{eqnewpsi2T}), where $q_{k,1}$ and $q_{k,2}$ in (\ref{eqPsik1}) and (\ref{eqPsik2}) can be simplified as
{\begingroup\makeatletter\def\f@size{10}\check@mathfonts
\def\maketag@@@#1{\hbox{\m@th\normalsize\normalfont#1}}\setlength{\arraycolsep}{0.0em}
\begin{eqnarray}
% \nonumber to remove numbering (before each equation)
{q_{k,1}} &=&
{
\begin{cases}\label{eqscheme2qk1alpha}
\displaystyle \int\nolimits_0^1\cdots\int\nolimits_0^1 {\frac{\mathrm{d}t_1\cdots\mathrm{d}t_k}{\left(B_{\textsf{s},\textsf{m}}\left(\alpha,\alpha,1, \frac{\theta_{\textsf{s}}}{\sum\nolimits_{i=1}^{k}t_i^{-\frac{\alpha}{2}}}\right)\right)^{K}}}, &  k=1,\cdots,K-1, \\
  0, & k=K,
\end{cases}
 }\\
{q_{k,2}} &=&   \begin{cases}\label{eqscheme2qk2alpha}
\displaystyle\frac{1}{\left(B_{\textsf{s},\textsf{m}}(\alpha,\alpha,1,\theta_{\textsf{s}},1)\right)^{K}}, &  k=1,\\
\displaystyle \int\nolimits_0^1\cdots\int\nolimits_0^1 {\frac{\mathrm{d}t_1\cdots\mathrm{d}t_{k-1}}{\left(B_{\textsf{s},\textsf{m}}\left(\alpha,\alpha,1, \frac{\theta_{\textsf{s}}}{1+\sum\nolimits_{i=1}^{k-1}t_i^{-\frac{\alpha}{2}}}\right)\right)^{K}}}, &  k=2,\cdots,K.
                \end{cases}
\end{eqnarray}\setlength{\arraycolsep}{5pt}\endgroup}Here, $B_{x,y}(\alpha_{x},\alpha_{y},T,\theta,u)$ is given by (\ref{eqscheme1Bxy}).
\end{Corollary}

\indent\indent \emph{Proof}: Corollary \ref{corollaryScheme2} can be proved in a similar way to Corollary \ref{corollaryScheme1K1}. We omit the details due to page limitation.

{In {Corollary} \ref{corollaryScheme2}, we obtain a closed-form expression of $\psi_{\mathrm{sch}_2}(\mathbf{T})$ when $K=1$ and $\alpha_{\textsf{s}}=\alpha_{\textsf{m}}=\alpha$. Furthermore, when $K\ge2$ and $\alpha_{\textsf{s}}=\alpha_{\textsf{m}}=\alpha$, we obtain a simpler expression of $\psi_{\mathrm{sch}_2}(\mathbf{T})$ than that of Theorem~2, which can be used to facilitate the numerical evaluation of $\psi_{\mathrm{sch}_2}(\mathbf{T})$.}

\begin{figure}[!t]
    \centering
    \subfloat{\includegraphics[width=2.5in]{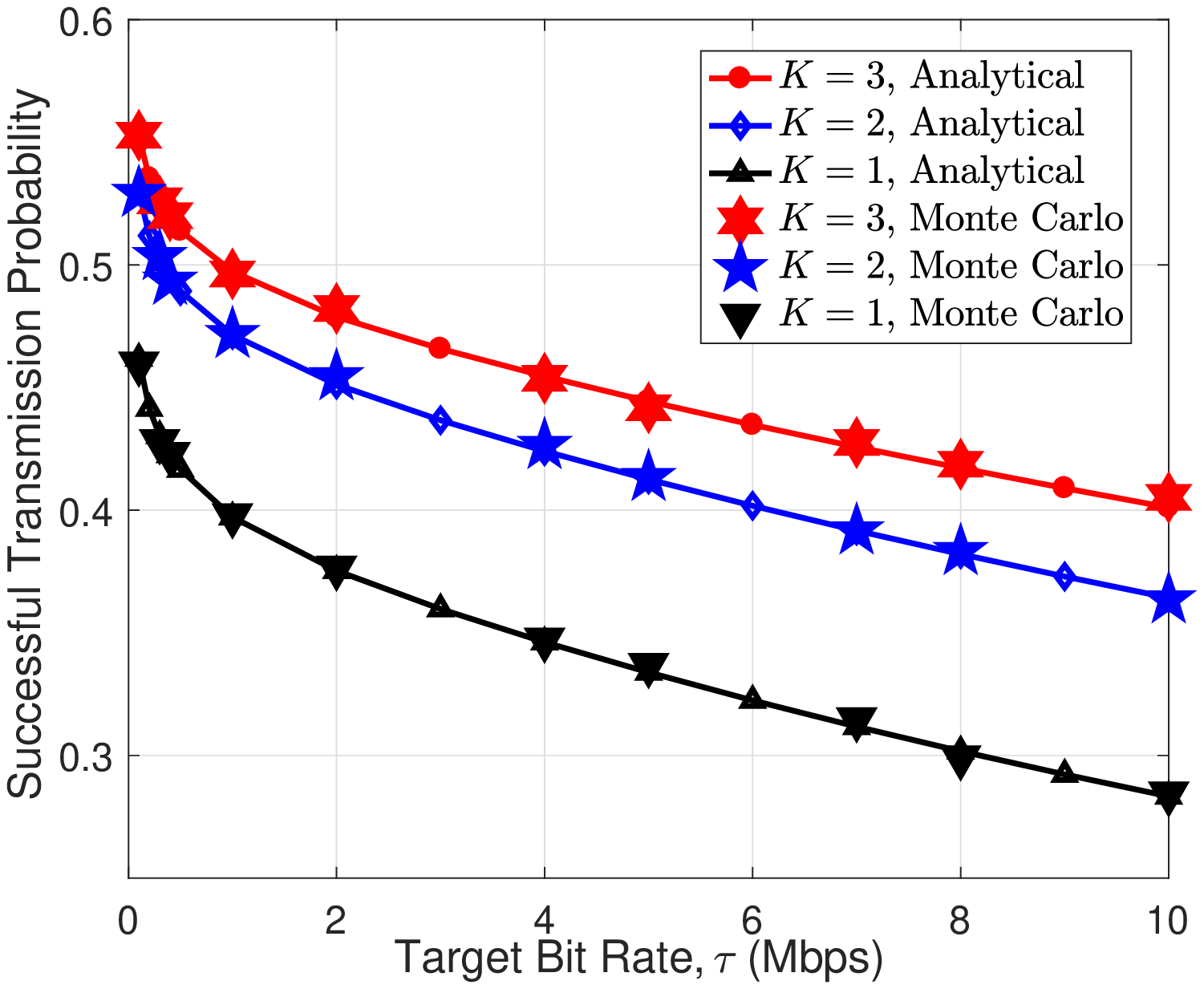}}
  \caption{STP ${\psi_{\mathrm{sch}_2}(\mathbf{T})}$ versus $\tau$. $M=2$, $N=10$, $T_1=0.9,T_2=0.8,T_3=0.3$, $T_n=0$ for $n=4,5,\cdots,N$, $\lambda_{\textsf{m}}=\frac{1}{ 500^2\pi}$ $\mathrm{m}^{-2}$, $\lambda_{\textsf{s}}=\frac{1}{ 50^2\pi}$ $\mathrm{m}^{-2}$, $P_{\textsf{m}}=43$ dBm, $P_{\textsf{s}}=23$ dBm, $\alpha_{\textsf{s}}=\alpha_{\textsf{m}}=4$, $W_{\textsf{m}}=0.2$ MHz, $W_{\textsf{s}}=20$ MHz, and $a_n=\frac{n^{-\gamma}}{\sum_{n \in\mathcal{N}}n^{-\gamma }}$ with Zipf exponent $\gamma=0.8$. In the Monte Carlo simulations, we choose a large spatial window, which is a square of $10^4 \times 10^4 $ $\mathrm{m}^2$, and the final simulation results are obtained by averaging over $10^4$ independent realizations.}\label{figpsi2}
\end{figure}

Fig. \ref{figpsi2} plots the STP $\psi_{\mathrm{sch}_2}(\mathbf{T})$ versus the target bit rate $\tau$ and verifies Theorem \ref{theorem2}. As expected, we see that the STP $\psi_{\mathrm{sch}_2}(\mathbf{T})$ decreases with the target bit rate $\tau$ and increases with the number of cooperative SBSs $K$. Besides, the marginal STP gain of including one more SBS into cooperation transmission decreases with $K$.
\subsection{Optimization of Successful Transmission Probability}
The caching distribution $\mathbf{T}$ significantly affects the STP under Scheme 2. We would like to maximize $\psi_{\mathrm{sch}_2}(\mathbf{T})$ in (\ref{eqnewpsi2T}) by optimizing $\mathbf{T}$. Note that, when studying Scheme 2, we focus on the region in which $\psi_{\textsf{s}_2,1}>\psi_{\textsf{m}}$. In this region, $u_0$ prefers receiving files from SBSs, as SBSs can offer a higher STP than the nearest MBS. Specifically, we have the following~problem.
\begin{Problem}[Optimization of STP under Scheme 2]\label{problem2}
\begin{eqnarray}
\psi_{\mathrm{sch}_2}^* \triangleq\mathop {\max }\limits_{\mathbf{T}} \;\;{\psi_{\mathrm{sch}_2}(\mathbf{T})}\label{p2}\\
s.t.\;\;\;\;(\ref{Tncons1}), (\ref{Tncons2}).\nonumber
\end{eqnarray}Here, $\mathbf{T}^*$ denotes the optimal solution, $\psi_{\mathrm{sch}_2}^*=\psi_{\mathrm{sch}_2}(\mathbf{T}^*)$ denotes the optimal value, and $\psi_{\mathrm{sch}_2}(\mathbf{T})$ is given by (\ref{eqnewpsi2T}).
\end{Problem}

{\color{black}Note that, different from $\psi_{\mathrm{sch}_1}(\mathbf{T})$ in (\ref{eqpsi1T}), $\psi_{\mathrm{sch}_2}(\mathbf{T})$ is a differentiable function of $\mathbf{T}$. Using KKT conditions,} we obtain the following optimality properties of {Problem}~\ref{problem2}.

\begin{Lemma}[Optimality Properties of Problem \ref{problem2}]\label{lemmaOptimalityPropertyeqproblem2Scheme2}
If $T_n^*$ is an optimal solution to Problem \ref{problem2}, then there exists $\nu\in \mathbbm{R}$ such that
{\begingroup\makeatletter\def\f@size{10}\check@mathfonts
\def\maketag@@@#1{\hbox{\m@th\normalsize\normalfont#1}}\setlength{\arraycolsep}{0.0em}
\begin{eqnarray}\label{eqscheme2optprop}
\begin{cases}
a_n\psi_{\textsf{ms}}^{'}(T_n^*)\le\nu,    & \mbox{if } T_n^*=0,\\
a_n\psi_{\textsf{ms}}^{'}(T_n^*)=  \nu,    & \mbox{if } T_n^*\in(0,1), \\
a_n\psi_{\textsf{ms}}^{'}(T_n^*)\ge\nu,    & \mbox{if } T_n^*=1,
\end{cases}
\end{eqnarray}\setlength{\arraycolsep}{5pt}\endgroup}where $\psi_{\textsf{ms}}^{'}(x)\triangleq \frac{\mathrm{d}\psi_{\textsf{ms}}(x)}{\mathrm{d}x}$ with $\psi_{\textsf{ms}}(x)$ given by (\ref{eqscheme2psims}). Furthermore, we have $1 \ge T_1^*\ge T_2^*\ge\cdots\ge T_{N}^*\ge 0$.
\end{Lemma}

\indent\indent{\emph{Proof}}: See Appendix D.

From {Lemma} \ref{lemmaOptimalityPropertyeqproblem2Scheme2}, we see that a file of higher popularity {should be stored} at the SBS tier with a higher probability (i.e., stored at more SBSs). In addition, we know that all files satisfying $T_n^*\in(0,1)$ must have the same $a_n\psi_{\textsf{ms}}^{'}(T_n^*)$, which is less than or equal to $a_n\psi_{\textsf{ms}}^{'}(T_n^*)$ for the files not stored at the SBS tier and greater than or equal to $a_n\psi_{\textsf{ms}}^{'}(T_n^*)$ for the files always being stored at the SBS tier.
\begin{algorithm}[!t]
\caption{\small Optimal Solution to the {Problem} \ref{problem2}}
\begin{algorithmic}[1]\label{algorithm2}\small
\IF {{$K=1$ or $\psi_{\textsf{s}_2,k+1}-\psi_{\textsf{s}_2,k}\le\psi_{\textsf{s}_2,k}-\psi_{\textsf{s}_2,k-1}$ for all $k=1,\cdots,K-1$}}
\STATE $f\leftarrow1$, $\nu_{\mathrm{lb}}= \min\limits_{n\in\mathcal{N}}\min\limits_{x\in[0,1]}a_n{\psi_{\textsf{ms}} ^{'}(x)} $, $\nu_{\mathrm{ub}}= \max\limits_{n\in\mathcal{N}}\max\limits_{x\in[0,1]}a_n{\psi_{\textsf{ms}} ^{'}(x)} $
\WHILE {$f=1$}
\STATE $\nu\leftarrow \nu_{\mathrm{lb}}+\frac{\nu_{\mathrm{ub}}-\nu_{\mathrm{lb}}}{2}$
\STATE Calculate $T_n^*$ using Lemma \ref{lemmaOptimalityPropertyeqproblem2Scheme2}, $n=1,\cdots,N$.
\IF {$\sum\nolimits_{n=1}^{N}T_n^*=M$}
\STATE $f\leftarrow0$
\ELSIF {$\sum\nolimits_{n=1}^{N}T_n^*>M$}\label{alg2step}
\STATE $\nu_{\mathrm{lb}}\leftarrow \nu$
\ELSE
\STATE $\nu_{\mathrm{ub}}\leftarrow \nu$
\ENDIF
\ENDWHILE
\ELSE
\STATE Calculate $T_n^*$ using gradient projection method, $n=1,\cdots,N$.
\ENDIF
\end{algorithmic}
\end{algorithm}

When $\psi_{\textsf{s}_2,k+1}-\psi_{\textsf{s}_2,k}\le\psi_{\textsf{s}_2,k}-\psi_{\textsf{s}_2,k-1}$ for all $k=1,\cdots,K-1$, from Remark \ref{propTh2Scheme2}, we know that $\psi_{\textsf{ms}}(T_n)$ is a concave function of $T_n$, and thus, $\psi_{\mathrm{sch}_2}(\mathbf{T})$ is a concave function of $\mathbf{T}$. In this case, Problem \ref{problem2} reduces to a convex optimization problem. In particular, from Remark \ref{propTh2Scheme2}, we know that when $K=1$, $\psi_{\textsf{ms}}(T_n)$ is a linear function of $T_n$, and thus, $\psi_{\mathrm{sch}_2}(\mathbf{T})$ is a linear function of $\mathbf{T}$. In this case, Problem \ref{problem2} reduces to a linear programming  problem. Thus, in these two cases, global optimal solutions to Problem \ref{problem2} can be obtained. In addition, $\psi_{\textsf{ms}}^{'}(T_n)$ is a monotonic decreasing function of $T_n$ when $\psi_{\textsf{s}_2,k+1}-\psi_{\textsf{s}_2,k}\le\psi_{\textsf{s}_2,k}-\psi_{\textsf{s}_2,k-1}$ for all $k=1,\cdots,K-1$ and is a constant when $K=1$. Thus, in these two cases, $\nu$ in Lemma \ref{lemmaOptimalityPropertyeqproblem2Scheme2} can be easily obtained by the bisection method, and $\mathbf{T}^*$ can be then determined by (\ref{eqscheme2optprop}). Therefore, in these two cases, we can use Lemma \ref{lemmaOptimalityPropertyeqproblem2Scheme2} to obtain global optimal solutions to Problem \ref{problem2}.

When $\psi_{\textsf{s}_2,k+1}-\psi_{\textsf{s}_2,k}>\psi_{\textsf{s}_2,k}-\psi_{\textsf{s}_2,k-1}$ for all $k=1,\cdots,K-1$, it is difficult to determine the concavity of $\psi_{\textsf{ms}}(T_n)$. In this case, a local optimal solution to Problem \ref{problem2} can be obtained using gradient projection method \cite{CuiCacheWirelessSignleTier}.
%Due to the multiple integrals in $\psi_{\textsf{ms}}(T_n)$, it is of huge complexity to optimize $\psi_{\mathrm{sch}_2}(\mathbf{T})$, when $K$ is large. From {Lemma} \ref{lemma1upperboundforScheme2}, we know that when $K\ge2$ and $\alpha_{\textsf{s}}=\alpha_{\textsf{m}}=\alpha$, there exists a simpler approximate expression $\widetilde{\psi}_{ms}(T_n)$ with a single-fold integral only. Thus, we can consider the optimization of $\widetilde\psi_{\mathrm{sch}_2}(\mathbf{T})$, instead of $\psi_{\mathrm{sch}_2}(\mathbf{T})$ for $K\ge2$ and $\alpha_{\textsf{s}}=\alpha_{\textsf{m}}=\alpha$ as an approximation, for the purpose of reducing computation complexity. Note that the above-mentioned properties for $\psi_{\mathrm{sch}_2}(\mathbf{T})$ also hold for $\widetilde{\psi}_{\mathrm{sch}_2}({\mathbf{T}})$. Therefore, the optimization of $\widetilde{\psi}_{\mathrm{sch}_2}({\mathbf{T}})$ can be done in a similar manner.

Finally, we can obtain a global (when $K=1$ or $\psi_{\textsf{s}_2,k+1}-\psi_{\textsf{s}_2,k}\le\psi_{\textsf{s}_2,k}-\psi_{\textsf{s}_2,k-1}$ for all $k=1,\cdots,K-1$) or local (when $\psi_{\textsf{s}_2,k+1}-\psi_{\textsf{s}_2,k}>\psi_{\textsf{s}_2,k}-\psi_{\textsf{s}_2,k-1}$ for all $k=1,\cdots,K-1$) optimal solution to Problem \ref{problem2} as summarized in Algorithm~2.
\section{Numerical Results}
In this section, we first compare the two proposed cooperative transmission schemes under the optimal caching designs. Then, under each scheme, we compare the optimal caching design with three baseline caching designs, i.e., MPC \cite{Cache-enabledsmallcellnetworksmodelingandtradeoffs}, IIDC \cite{Bharath2015Caching} and UC \cite{Tamoorulhassan2015Modeling}. {Specifically, under Scheme 1, Algorithm \ref{algorithm1} is used to obtain a local or global optimal caching distribution to Problem \ref{problem1}.
%Note that, when $K\ge2$ and $\alpha_{\textsf{s}}=\alpha_{\textsf{m}}=\alpha$, we consider the optimization of $\widetilde{\psi}_{\mathrm{sch}_1}(\mathbf{T})$ in Lemma \ref{lemma1upperboundforScheme1}, instead of $\psi_{\mathrm{sch}_1}(\mathbf{T})$ in Theorem \ref{theorem1} as an approximation for the purpose of reducing computational complexity.
Under Scheme 2, Algorithm~2 is used to obtain a local or global optimal caching distribution to Problem \ref{problem2}.
%Note that, when $K\ge2$ and $\alpha_{\textsf{s}}=\alpha_{\textsf{m}}=\alpha$, we consider the optimization of $\widetilde\psi_{\mathrm{sch}_2}(\mathbf{T})$, instead of $\psi_{\mathrm{sch}_2}(\mathbf{T})$ as an approximation for the purpose of reducing computational complexity.
Unless otherwise stated, our simulation environment settings are as follows: $\tau=1$ Mbps, $\lambda_{\textsf{m}}=\frac{1}{ 500^2\pi}$ $\mathrm{m}^{-2}$, $\lambda_{\textsf{s}}=\frac{1}{ 50^2\pi}$ $\mathrm{m}^{-2}$, $P_{\textsf{m}}=43$ dBm, $P_{\textsf{s}}=23$ dBm, $W_{\textsf{m}}=0.2$ MHz, $W_{\textsf{s}}=20$ MHz, $\alpha_{\textsf{s}}=\alpha_{\textsf{m}}=4$, $N=100$ and $a_n=\frac{n^{-\gamma}}{\sum_{n \in\mathcal{N}}n^{-\gamma }}$, where $\gamma$ is the Zipf exponent.}
\subsection{Comparisons Between Scheme 1 and Scheme 2}
\begin{figure}[!t]
\centering
\subfloat[$\tau=1$ Mbps]{
\includegraphics[width=2.5in]{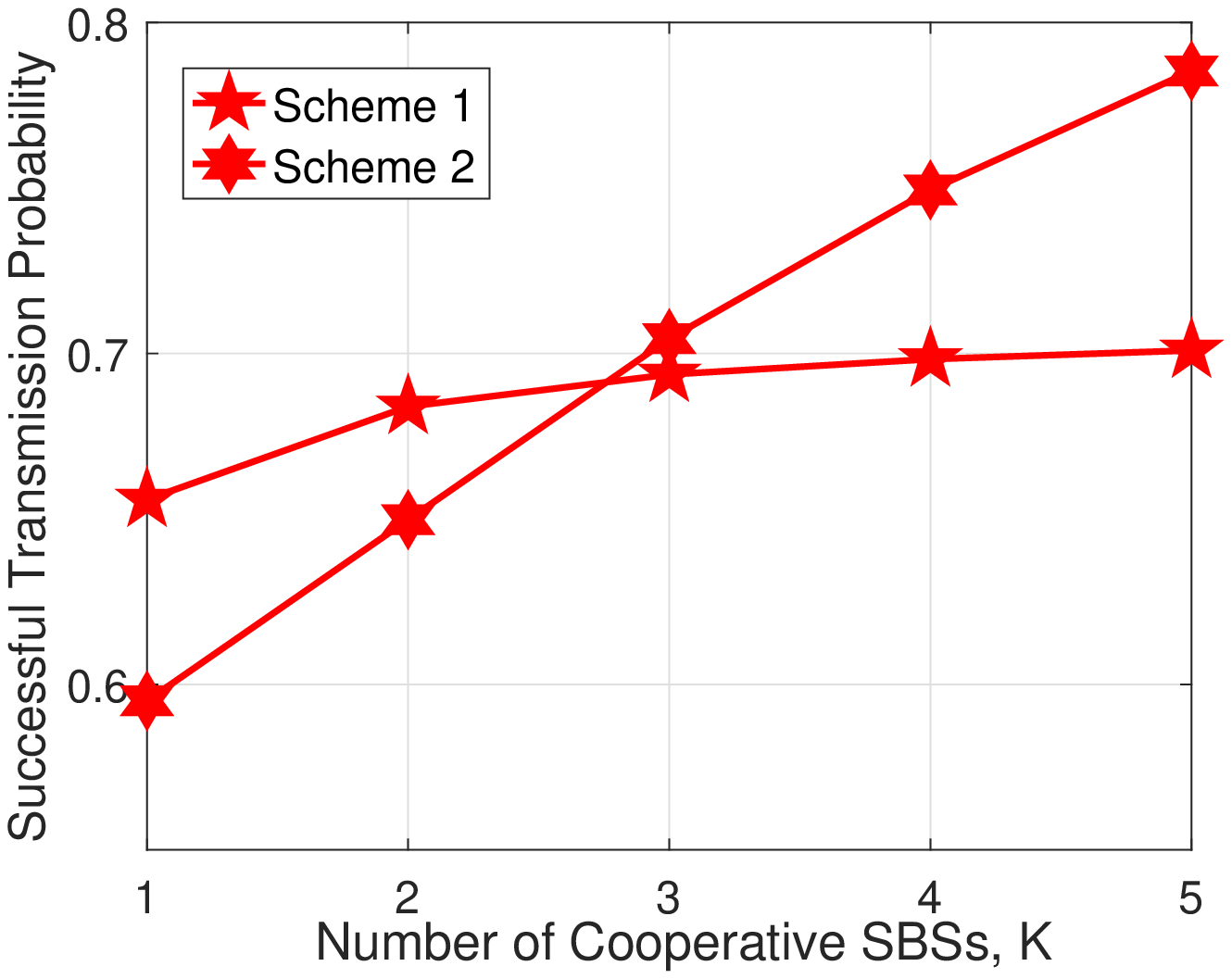}
\label{scheme1andscheme2comparisionoptimizationvsK}
}
\hfil
\subfloat[$K=4$]{
\includegraphics[width=2.5in]{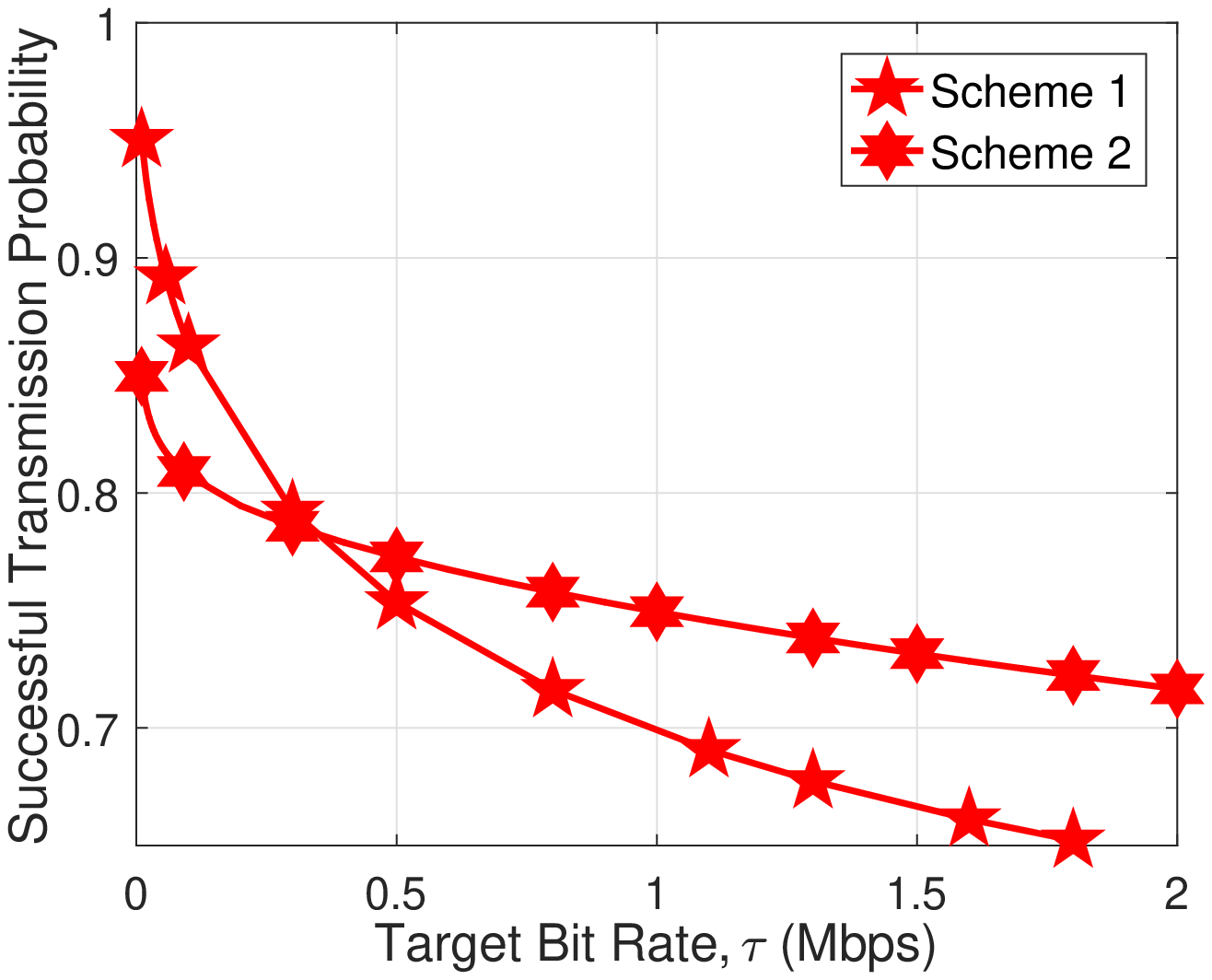}
\label{sch1vssch2theta}
}
\caption{Comparisons between Scheme 1 and Scheme 2 where $M=25$ and $\gamma=0.8$.}
\label{ComparisonBetweenScheme1andScheme2}
\end{figure}
In this part, we compare the two cooperative transmission schemes under the optimal caching designs. Fig. \ref{ComparisonBetweenScheme1andScheme2} illustrates the STP under each scheme versus the number of cooperative SBSs $K$ and the target bit rate $\tau$, respectively. From Fig. \ref{ComparisonBetweenScheme1andScheme2}(a), we observe that the STP under each scheme increases with $K$, since larger $K$ leads to higher desired signal power and lower interference power. In addition, when $K$ is large, e.g., $K\ge2$, the marginal STP increase w.r.t. $K$ under Scheme 1 becomes small. This is because the average desired signal power from an SBS far from the typical user is weak, and the advantage of incorporating it in the joint transmission is negligible. While, for all $K=1,\cdots,5$, the marginal STP increase w.r.t. $K$ under Scheme~2 is large. This is because when $K=1,\cdots,5$, the nearest $K$ SBSs are still close to the typical user, including one more SBS in the joint transmission greatly increases the desired signal power, and silencing one more SBS significantly reduces interference power. Furthermore, when $K$ is small, Scheme~1 outperforms Scheme~2, implying that including one more SBS in the joint transmission is {preferable} in this region; when $K$ is large, Scheme~2 outperforms Scheme~1, implying that silencing one more SBS is {preferable} in this region. From Fig. \ref{ComparisonBetweenScheme1andScheme2}(b), we observe that the STP under each scheme decreases with $\tau$. In addition, when $\tau$ is small, Scheme 1 outperforms Scheme 2, {implying that SBS joint transmission is preferable in this region. When $\tau$ is large, Scheme 2 outperforms Scheme~1, implying that SBS silencing is preferable in this~region.}
\subsection{Comparisons Between the Optimal Caching Designs and Baselines}
\begin{figure*}[!t]
\centering
\subfloat[Scheme 1]{
\includegraphics[width=2.5in]{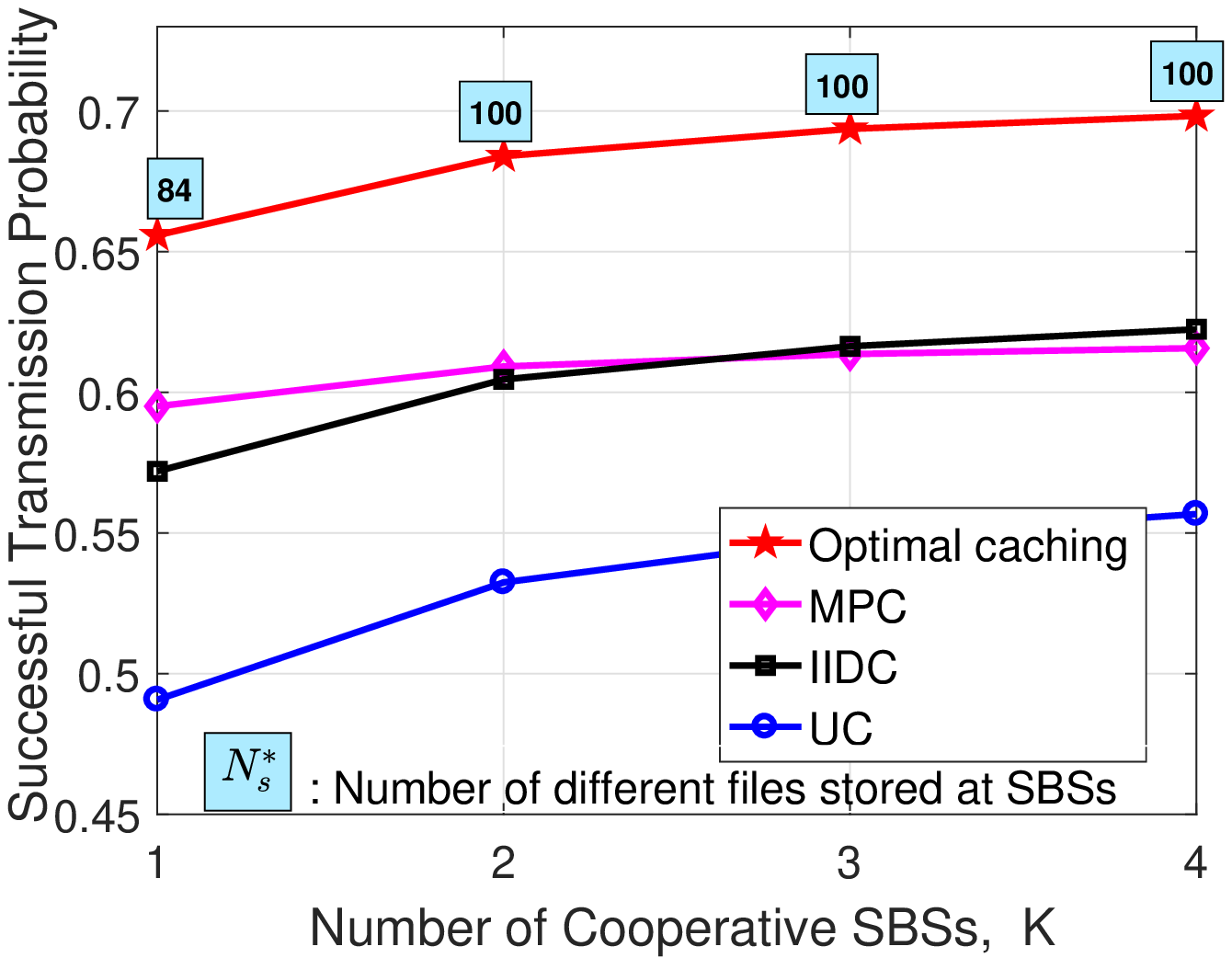}
}
\hfil
\subfloat[Scheme 2]{
\includegraphics[width=2.5in]{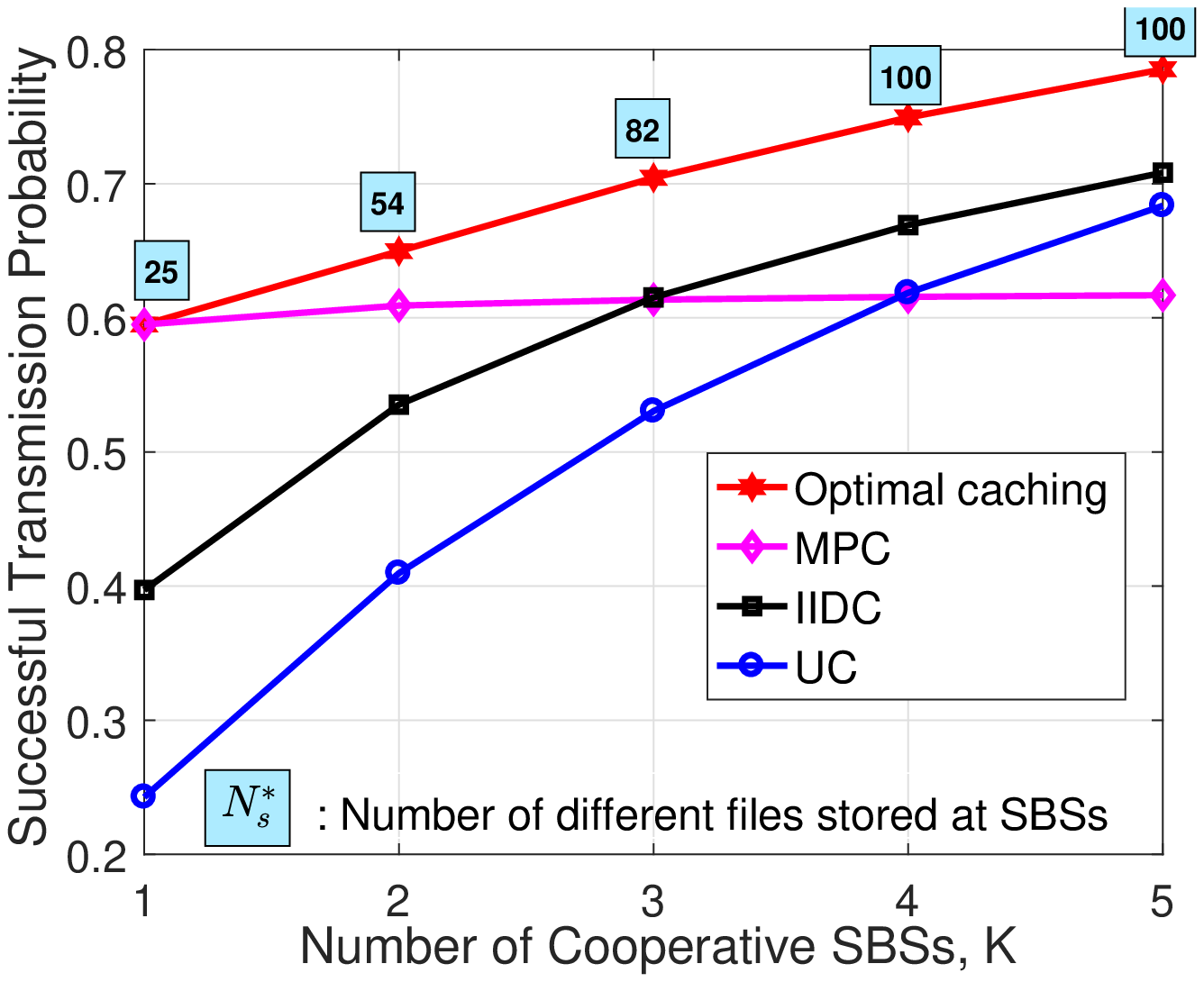}
}
\caption{Comparisons between the optimal caching and baselines under various number of cooperative SBSs $K$ at $M=25$ and $\gamma=0.8$.}
\label{stpvsK}
\end{figure*}
\begin{figure*}[!t]
\centering
\subfloat[Scheme 1]{
\includegraphics[width=2.5in]{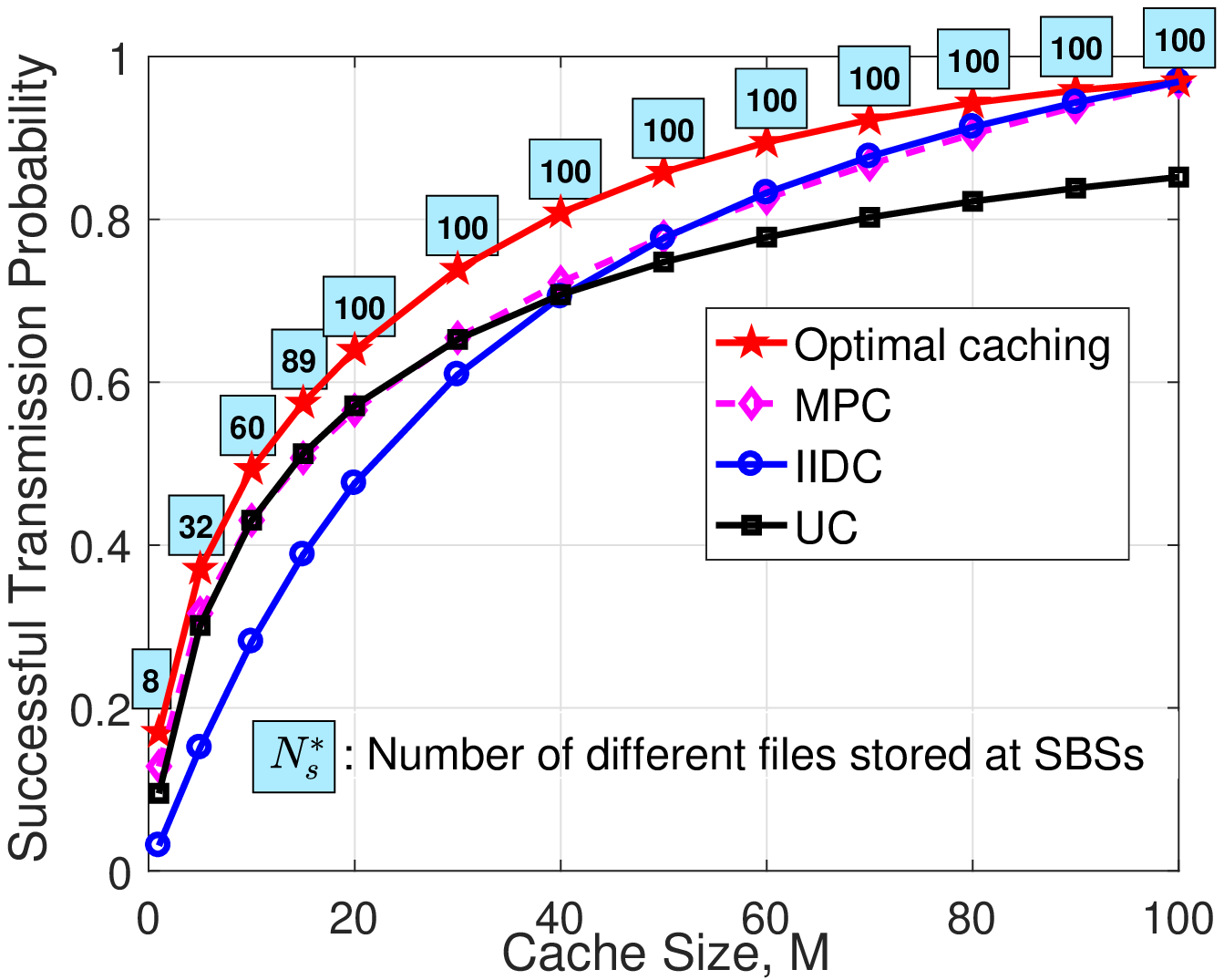}
}\hfil
\subfloat[Scheme 2]{
\includegraphics[width=2.5in]{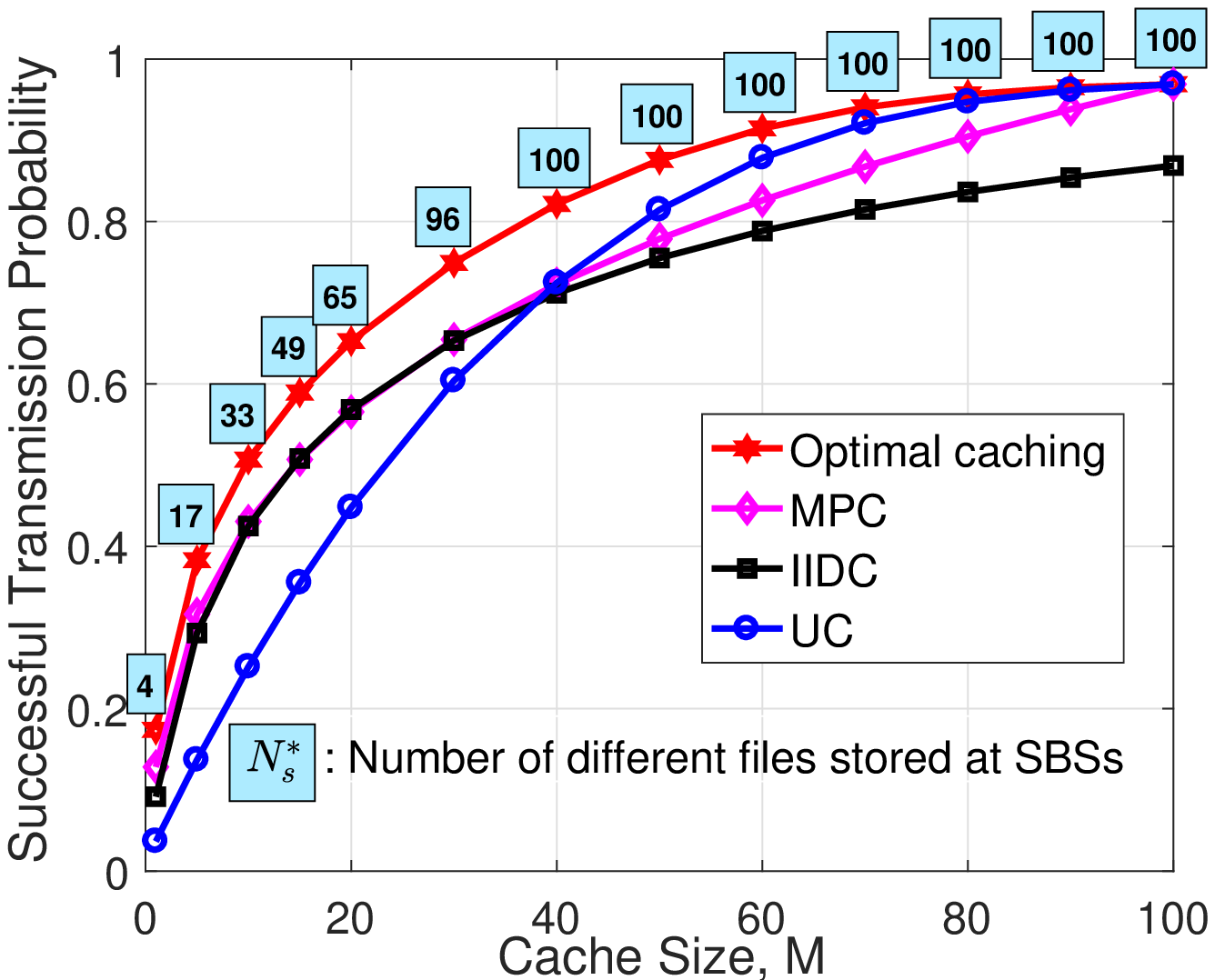}
}
\caption{Comparisons between the optimal caching and baselines under various cache size $M$ at $K=3$ and $\gamma=0.8$.}
\label{stpvsM}
\end{figure*}
\begin{figure*}[!t]
\centering
\subfloat[Scheme 1]{
\includegraphics[width=2.5in]{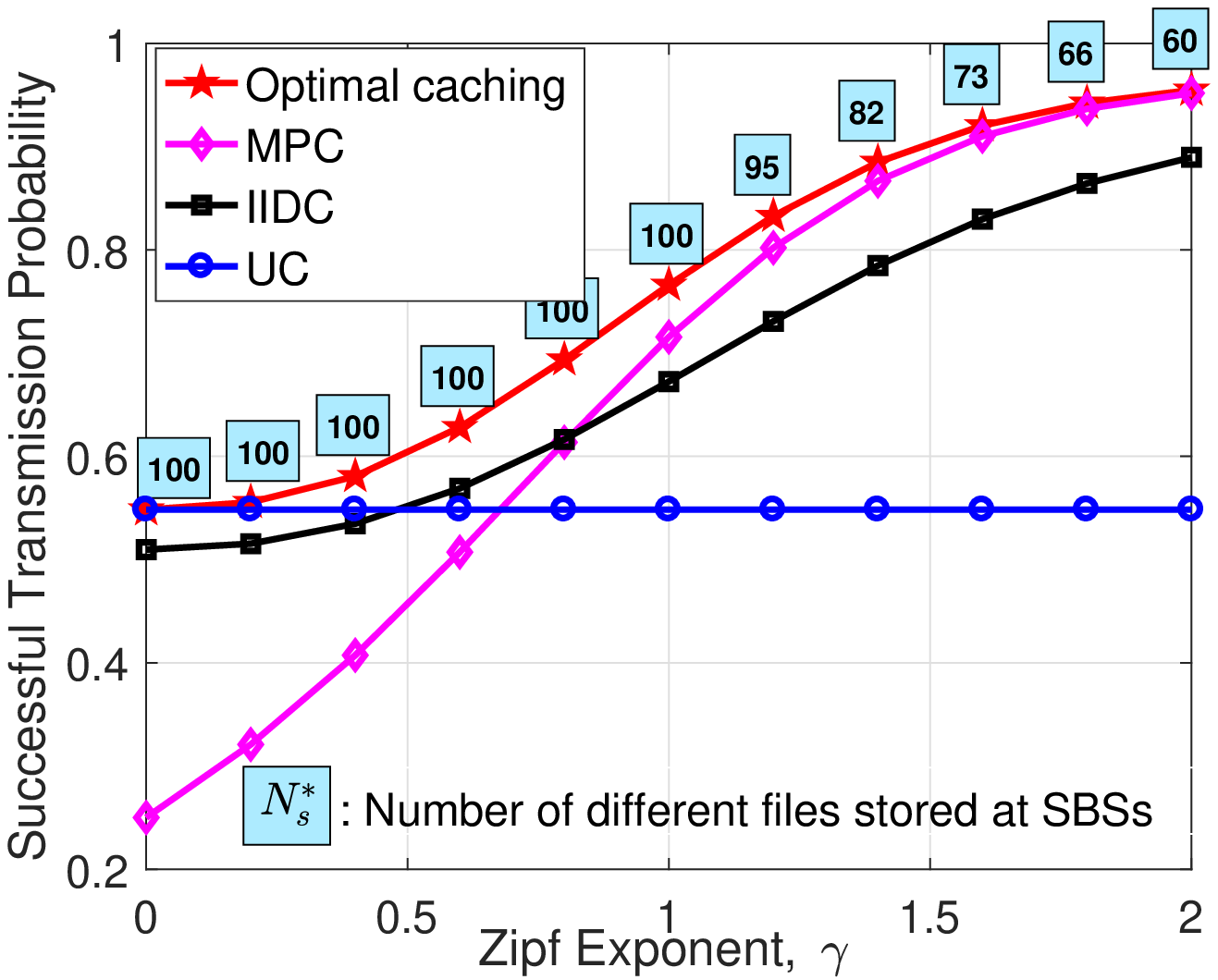}
}\hfil
\subfloat[Scheme 2]{
\includegraphics[width=2.5in]{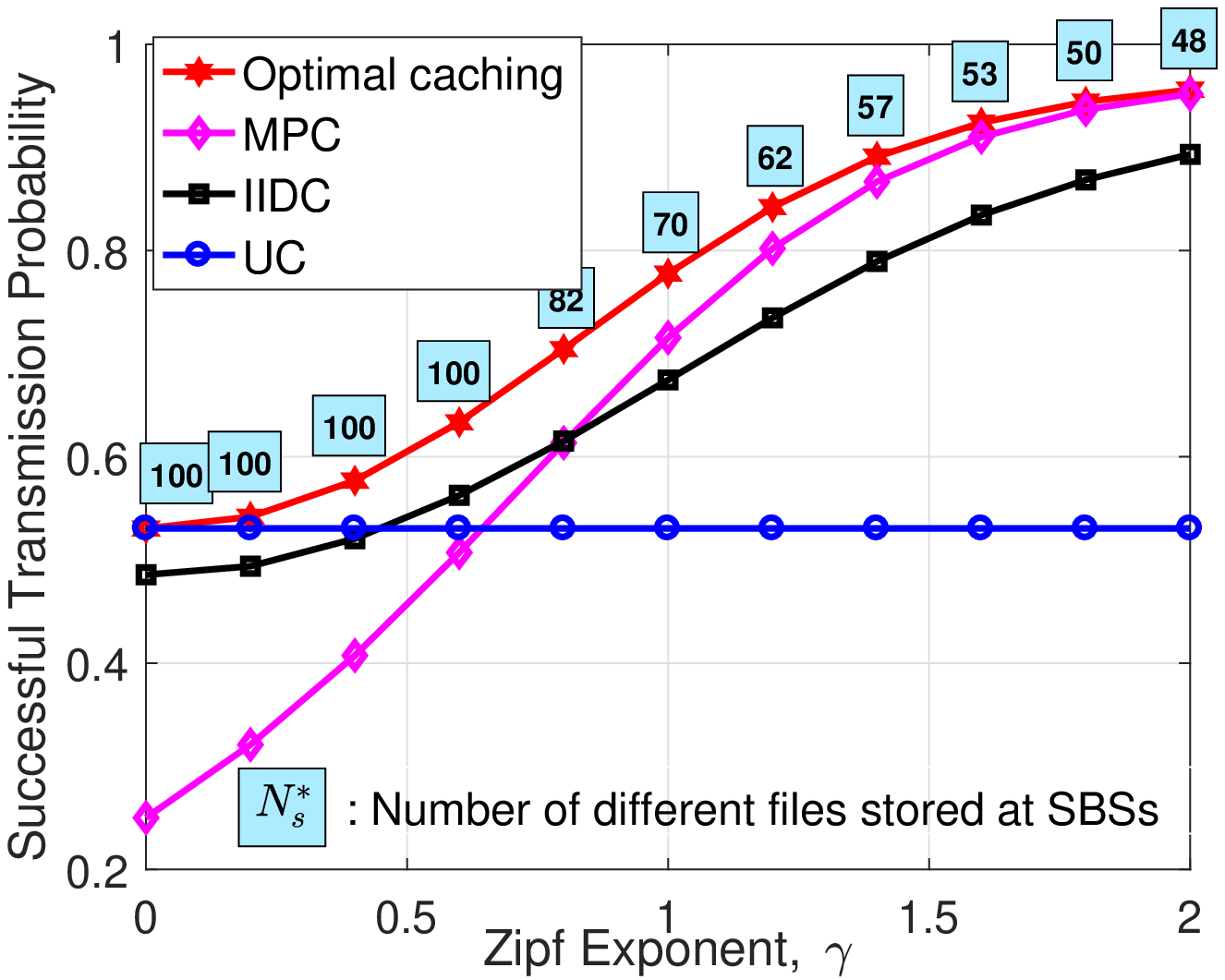}
}
\caption{Comparisons between the optimal caching and baselines under various Zipf exponent $\gamma$ at $K=3$ and $M=25$.}
\label{stpvsGamma}
\end{figure*}

In this part, under each cooperative transmission scheme, we compare the optimal caching design and three baseline caching designs. From Fig. \ref{stpvsK}--\ref{stpvsGamma}, we can observe that the optimal caching design outperforms all three baselines under each cooperation transmission scheme. In the following, similarly, for the the optimal caching design under Scheme~2, we denote $\max\{n\in\mathcal{N}:T_n^*\ge10^{-2}\}$ by $N_{\textsf{s}}^*$, indicating the number of different files stored at SBSs. Note that, here we choose $10^{-2}$ as the lower bound instead of $0$ to accommodate the calculation error by the numerical algorithm.

Specifically, Fig. \ref{stpvsK} illustrates the STP versus the number of cooperative SBSs $K$. From Fig. \ref{stpvsK}, we observe that under the optimal caching, a larger $K$ leads to a larger $N_{\textsf{s}}^*$ (up to $N$), which means that the larger the number of cooperative SBSs, the more files should be stored at the SBS tier. Since MPC stores only the $M\le N_{\textsf{s}}^*$ most popular files at each SBS tier, the optimal caching can achieve higher file diversity and thus outperforms MPC. As UC cannot exploit the file popularity to improve the performance, and IIDC may store multiple copies of the same file, leading to storage waste, the optimal caching outperforms UC and IIDC.

Fig. \ref{stpvsM} illustrates the STP versus the cache sizes $M$. We can see that with $M$ increasing, the performance of all designs increases. This is because as $M$ increases, each SBS can store more files, and the probability that a requested file is stored at the cooperative SBSs increases. Furthermore, we see that when $M$ increases, $N_{\textsf{s}}^*$ increases (up to $N$), implying that the larger the cache size, the more files will be stored at the SBS tier. In addition, when $M$ becomes sufficiently large, the STP gap between the optimal caching and MPC or UC becomes much~smaller.

Fig. \ref{stpvsGamma} illustrates the STP versus the Zipf exponent $\gamma$. We can see that the performance of the optimal caching, MPC and IIDC increases with $\gamma$, whereas the STP of UC stays flat with $\gamma$ since it does not exploit any file popularity. In addition, for a large $\gamma$, the optimal caching reduces to MPC, implying that only a small number of the most popular files should be stored at the SBS tier in this region. While for a small $\gamma$, the optimal caching reduces to UC, implying that a large number of files should be stored at the SBS tier in this region.
\section{Conclusion}
In this paper, we jointly considered SBS caching and cooperation in a downlink large-scale  HetNet. Based on a random caching design, two cooperative transmission schemes were proposed. Utilizing tools from stochastic geometry, we derived tractable expressions for the STP under each scheme. Then, under each scheme, we considered the STP maximization. By exploring optimality properties and using optimization techniques, a local optimal solution in the general case and global optimal solutions in some special cases were obtained for each scheme. Under each scheme, compared with some existing caching designs in the literature, e.g., MPC, IIDC and UC, the optimal caching design achieved better STP performance.

\appendices
\section*{Appendix A: Proof of the {Theorem} \ref{theorem1}}\label{ProofForTheorem1}
To prove Theorem \ref{theorem1}, we rewrite (\ref{psi1Def}) as follows:
{\begingroup\makeatletter\def\f@size{10}\check@mathfonts
\def\maketag@@@#1{\hbox{\m@th\normalsize\normalfont#1}}\setlength{\arraycolsep}{0.0em}
\begin{eqnarray}\label{eqApppsisch1}
\psi_{\mathrm{sch}_1}(\mathbf{T}) &=& \sum\limits_{n \in \mathcal{N}}{a_n}\bigg( \underbrace{\Pr \left[ {W_{\textsf{m}}{{\log }_2}\left( {1 + {\gamma _{\textsf{m}}}} \right) > \tau }\right]}_{\triangleq\psi_{\textsf{m}}}\mathbbm{1}\left[T_n=0\right]+\underbrace{\Pr \left[ {W_{\textsf{s}}{{\log }_2}\left( {1 + {\gamma _{\textsf{s}_1,n}}} \right) > \tau }\right]}_{\triangleq\psi_{\textsf{s}_1}(T_n)}\mathbbm{1}\left[T_n>0\right]\bigg),
\end{eqnarray}\setlength{\arraycolsep}{5pt}\endgroup}where $\gamma_{\textsf{m}}$ and $\gamma_{\textsf{s}_1,n}$ are given by (\ref{eqDefineSIRGammam}) and (\ref{eqgammas1n}), respectively. Based on (\ref{eqApppsisch1}), we calculate $\psi_{\textsf{m}}$ and $\psi_{\textsf{s}_1}(T_n)$, respectively.
\subsection*{Calculation of $\psi_{\textsf{m}}$}
Consider the case that $u_0$ is served by its nearest MBS. First, we rewrite the SIR expression in (\ref{eqDefineSIRGammam}) as $\gamma _{\textsf{m}}= \frac{X_{\textsf{m}}}{{I_{\textsf{s}} + I_{\textsf{m}}}}$, where $X_{\textsf{m}}\triangleq{{P_{\textsf{m}}}{h_{\textsf{m},l_{\textsf{m}}}}r_{\textsf{m},l_{\textsf{m}}}^{ - {\alpha_{\textsf{m}}}}}$, $I_{\textsf{s}}\triangleq\sum\nolimits_{{l} \in \Phi_{\textsf{s}}} {{P_{\textsf{s}}}{{\left| {{h_{\textsf{s},l}}} \right|}^2}r_{\textsf{s},l}^{ - {\alpha_{\textsf{s}}}}}$ and $I_{\textsf{m}}\triangleq\sum\nolimits_{{l} \in \Phi_{\textsf{m}}\setminus {l_{\textsf{m}}}} {{P_{\textsf{m}}}{{\left| {{h_{\textsf{m},l}}} \right|}^2}r_{\textsf{m},l}^{ - {\alpha_{\textsf{m}}}}}$. Conditioning on $r_{\textsf{m},l_{\textsf{m}}}=r$, we have:
{\begingroup\makeatletter\def\f@size{10}\check@mathfonts
\def\maketag@@@#1{\hbox{\m@th\normalsize\normalfont#1}}\setlength{\arraycolsep}{0.0em}
\begin{eqnarray}\label{eqAppdPsim}
\psi_{\textsf{m},r_{\textsf{m},l_{\textsf{m}}}}(r)&\triangleq& \mathrm{Pr}\left[W_{\textsf{m}}\log_2(1+\gamma_{\textsf{m}})>\tau|r_{\textsf{m},l_{\textsf{m}}}=r\right]\nonumber\\
&=&\mathrm{E}_{I_{\textsf{s}},I_{\textsf{m}}}\left[\mathrm{Pr}\left[X_{\textsf{m}}\ge\theta_{\textsf{m}}(I_{\textsf{s}}+I_{\textsf{m}})|r_{\textsf{m},l_{\textsf{m}}}=r\right]\right]\mathop=\limits^{(\mathrm{a})}\mathrm{E}_{I_{\textsf{s}},I_{\textsf{m}}}\left[\exp\left(-\frac{r^{\alpha_{\textsf{m}}}}{P_{\textsf{m}}}\theta_{\textsf{m}}(I_{\textsf{s}}+I_{\textsf{m}})\right)\right]\nonumber\\
&\mathop=\limits^{(\mathrm{b})}&\underbrace{\mathrm{E}_{I_{\textsf{s}}}\left[\exp\left(-\frac{r^{\alpha_{\textsf{m}}}}{P_{\textsf{m}}}\theta_{\textsf{m}} I_{\textsf{s}}\right)\right]}_{\triangleq \mathcal{L}_{I_{\textsf{s}}}(z,r)|_{z=\frac{r^{\alpha_{\textsf{m}}}}{P_{\textsf{m}}}\theta_{\textsf{m}}}}
\underbrace{\mathrm{E}_{I_{\textsf{m}}}\left[\exp\left(-\frac{r^{\alpha_{\textsf{m}}}}{P_{\textsf{m}}}\theta_{\textsf{m}} I_{\textsf{m}}\right)\right]}_{\triangleq \mathcal{L}_{I_{\textsf{m}}}(z,r)|_{z=\frac{r^{\alpha_{\textsf{m}}}}{P_{\textsf{m}}}\theta_{\textsf{m}}}},
\end{eqnarray}\setlength{\arraycolsep}{5pt}\endgroup}where $\theta_{\textsf{m}}\triangleq2^{\tau/W_{\textsf{m}}}-1$, (a) is obtained by noting that $X_{\textsf{m}}$ is an exponential random variable with mean $P_{\textsf{m}}r^{-\alpha_{\textsf{m}}}$, i.e., $X_{\textsf{m}}\thicksim \exp\left({r^{\alpha_{\textsf{m}}}}{\theta_{\textsf{m}} P_{\textsf{m}}^{-1} }\right)$, and (b) is due to the independence of the Rayleigh fading channels and the independence of the homogeneous PPPs. $\mathcal{L}_{I_{\textsf{s}}}(z,r)$ and $\mathcal{L}_{I_{\textsf{m}}}(z,r)$ represent the Laplace transforms of the interference $I_{\textsf{s}}$ and $I_{\textsf{m}}$, respectively. To calculate $\psi_{\textsf{m},r_{\textsf{m},l_{\textsf{m}}}}(r)$ according to (\ref{eqAppdPsim}), we first calculate $\mathcal{L}_{I_{\textsf{s}}}(z,r)$ and $\mathcal{L}_{I_{\textsf{m}}}(z,r)$, respectively, as follows:
{\begingroup\makeatletter\def\f@size{10}\check@mathfonts
\def\maketag@@@#1{\hbox{\m@th\normalsize\normalfont#1}}\setlength{\arraycolsep}{0.0em}
\begin{eqnarray}\label{eqAppdPsimLIs}
\mathcal{L}_{I_{\textsf{s}}}(z,r) &=& \mathrm{E}\left[\exp\left(-z\sum\limits_{{l} \in \Phi_{\textsf{s}}} {{P_{\textsf{s}}}{{\left| {{h_{\textsf{s},l}}} \right|}^2}r_{\textsf{s},l}^{ - {\alpha_{\textsf{s}}}}}\right)\right]= \mathrm{E}\left[\prod\limits_{{l} \in \Phi_{\textsf{s}}}\exp\left(-z {{P_{\textsf{s}}}{{\left| {{h_{\textsf{s},l}}} \right|}^2}r_{\textsf{s},l}^{ - {\alpha_{\textsf{s}}}}}\right)\right]\nonumber\\
&\mathop=\limits^{(\mathrm{c})}&\exp\left(-2\pi\lambda_{\textsf{s}}\int_0^\infty\left(1-\frac{1}{1+z P_{\textsf{s}}v^{-\alpha_{\textsf{s}}}}\right)v\mathrm{d}v\right)=\exp\left(-{\frac{2\pi^2}{\alpha_{\textsf{s}}}\csc\left(\frac{2\pi}{\alpha_{\textsf{s}}}\right) \lambda_{\textsf{s}}}\left(z P_{\textsf{s}}\right)^{{2}/{\alpha_{\textsf{s}}}}\right),\\
\mathcal{L}_{I_{\textsf{m}}}(z,r)&=&\mathrm{E}\left[\exp\left(-z\sum\limits_{{l} \in \Phi_{\textsf{m}}\setminus{l_{\textsf{m}}}} {{P_{\textsf{m}}}{{\left| {{h_{\textsf{m},l}}} \right|}^2}r_{\textsf{m},l}^{ - {\alpha_{\textsf{m}}}}}\right)\right]= \mathrm{E}\left[\prod\limits_{l\in\Phi_{\textsf{m}}\setminus{l_{\textsf{m}}}}\exp\left(-z {{P_{\textsf{m}}}{{\left| {{h_{\textsf{m},l}}} \right|}^2}r_{\textsf{m},l}^{ - {\alpha_{\textsf{m}}}}}\right)\right] \label{eqAppdPsimLIm}\nonumber \\
&\mathop=\limits^{(\mathrm{d})}&\exp\left(-2\pi\lambda_{\textsf{m}}\int_r^\infty\left(1-\frac{1}{1+z P_{\textsf{m}}v^{-\alpha_{\textsf{m}}}}\right)v\mathrm{d}v\right)\nonumber\\
&=&\exp\left(-\pi\lambda_{\textsf{m}}r^2 \left({_2F_1}\left(-\frac{2}{\alpha_{\textsf{m}}},1;1-\frac{2}{\alpha_{\textsf{m}}},-\frac{z P_{\textsf{m}}}{r^{\alpha_{\textsf{m}}}}\right)-1\right)\right),
\end{eqnarray}\setlength{\arraycolsep}{5pt}\endgroup}where (c) and (d) are obtained by utilizing the probability generating functional of PPP \cite{Haenggi2012Stochastic}.
Substituting (\ref{eqAppdPsimLIs}) and (\ref{eqAppdPsimLIm}) into (\ref{eqAppdPsim}), we obtain $\psi_{\textsf{m},r_{\textsf{m},l_{\textsf{m}}}}(r)$ as follows:
{\begingroup\makeatletter\def\f@size{10}\check@mathfonts
\def\maketag@@@#1{\hbox{\m@th\normalsize\normalfont#1}}\setlength{\arraycolsep}{0.0em}
\begin{eqnarray}\label{eqAppdPsim2}
\psi_{\textsf{m},r_{\textsf{m},l_{\textsf{m}}}}(r) &=& \exp\left(-{\frac{2\pi^2}{\alpha_{\textsf{s}}}\csc\left(\frac{2\pi}{\alpha_{\textsf{s}}}\right) \lambda_{\textsf{s}}}\left(\frac{\theta_{\textsf{m}} P_{\textsf{s}}}{P_{\textsf{m}}}\right)^{{2}/{\alpha_{\textsf{s}}}}
r^{{2\alpha_{\textsf{m}}}/{\alpha_{\textsf{s}}}}\right)\nonumber\\
&&{\times}\: \exp\left(-\pi\lambda_{\textsf{m}}r^2\left({_2F_1}\left(-\frac{2}{\alpha_{\textsf{m}}},1;1-\frac{2}{\alpha_{\textsf{m}}},-\theta_{\textsf{m}}\right)-1\right)\right).
\end{eqnarray}\setlength{\arraycolsep}{5pt}\endgroup

Now, we calculate $\psi_{\textsf{m}}$ by first removing the condition of $\psi_{\textsf{m},r_{\textsf{m},l_{\textsf{m}}}}(r)$ on $r_{\textsf{m},l_{\textsf{m}}}=r$. By noting that the p.d.f. of $r_{\textsf{m},l_{\textsf{m}}}$ is  $f_{r_{\textsf{m},l_{\textsf{m}}}}(r)=2\pi\lambda_{\textsf{m}}r\exp(-\pi\lambda_{\textsf{m}}r^2)$ \cite{10}, we have:
{\begingroup\makeatletter\def\f@size{10}\check@mathfonts
\def\maketag@@@#1{\hbox{\m@th\normalsize\normalfont#1}}\setlength{\arraycolsep}{0.0em}
\begin{eqnarray*}
\psi_{\textsf{m}}&=&\int_0^\infty\psi_{\textsf{m},r_{\textsf{m},l_{\textsf{m}}}}(r)f_{r_{\textsf{m},l_{\textsf{m}}}}(r)\mathrm{d}r.
\end{eqnarray*}\setlength{\arraycolsep}{5pt}\endgroup}By using the change of variable $u=\pi\lambda_{\textsf{m}}r^2$ and using the definition of $B_{x,y}(\alpha_{x},\alpha_{y},T,\theta,u)$, we can get $\psi_{\textsf{m}}$ in Theorem \ref{theorem1}.
\subsection*{Calculation of $\psi_{\textsf{s}_1}(T_n)$}
Consider the case that $u_0$ requesting file $n$ is jointly served by the SBSs in $\mathcal{C}_{1,n}$. There are three types of interferers, namely, i) all the other SBSs storing file $n$ besides the SBSs in $\mathcal{C}_{1,n}$, ii) all the SBSs not storing file $n$, and iii) all the MBSs. Thus, we rewrite the SIR expression in (\ref{eqgammas1n}) as $\gamma _{\textsf{s}_1,n} = \frac{X_{\textsf{s}_1}}{I_{\textsf{s},n}+I_{\textsf{s}_,-n} + I_{\textsf{m}}}$, where $X_{\textsf{s}_1}\triangleq {\left| {\sum\nolimits_{{l_{\textsf{s}}} \in {\mathcal{C}_{1,n}}} {\sqrt {{P_{\textsf{s}}}} {h_{\textsf{s},l_{\textsf{s}}}}r_{\textsf{s},l_{\textsf{s}}}^{ - {\alpha_{\textsf{s}}}/2}} } \right|^2}$, $I_{\textsf{s},n}\triangleq \sum\nolimits_{l \in {\Phi_{\textsf{s},n}\backslash \mathcal{C}_{1,n}}} {{P_{\textsf{s}}}{{\left| {{h_{\textsf{s},l}}} \right|}^2}r_{\textsf{s},l}^{ - {\alpha_{\textsf{s}}}}}$, $I_{\textsf{s},-n}\triangleq \sum\nolimits_{l \in {\Phi_{\textsf{s},-n}}} {{P_{\textsf{s}}}{{\left| {{h_{\textsf{s},l}}} \right|}^2}r_{\textsf{s},l}^{ - {\alpha_{\textsf{s}}}}}$ and $I_{\textsf{m}}\triangleq\sum\nolimits_{{l} \in \Phi_{\textsf{m}}} {{P_{\textsf{m}}}{{\left| {{h_{\textsf{m},l}}} \right|}^2}r_{\textsf{m},l}^{ - {\alpha_{\textsf{m}}}}}$ with $\Phi_{\textsf{s},-n}\triangleq\Phi_{\textsf{s}}\setminus\Phi_{\textsf{s},n}$ denoting the homogeneous PPP with density $(1-T_n)\lambda_{\textsf{s}}$ generated by SBSs not storing file $n$.

For notation simplicity, we denote by $X_1,\cdots,X_K$ the distances between the $K$ SBSs in $\mathcal{C}_{1,n}$ and $u_0$, where $X_K$ particularly denotes the distance between the $K$th nearest SBS in $\mathcal{C}_{1,n}$ and $u_0$, i.e., $0< X_k\le X_K$ for $k=1,\cdots,K-1$. Denote $\mathbf{X}\triangleq(X_k)_{k=1,\cdots,K}$. Conditioning on $\mathbf{X}=\mathbf{x}$, where $\mathbf{x}\triangleq(x_k)_{k=1,\cdots,K}$, we~have:
{\begingroup\makeatletter\def\f@size{10}\check@mathfonts
\def\maketag@@@#1{\hbox{\m@th\normalsize\normalfont#1}}\setlength{\arraycolsep}{0.0em}
\begin{eqnarray}\label{eqAppdPsis1}
\psi_{\textsf{s}_1,\mathbf{X}}(T_n,\mathbf{x})&\triangleq& \mathrm{Pr}\left[W_{\textsf{s}}\log_2(1+\gamma_{\textsf{s}_1,n})>\tau|\mathbf{X}=\mathbf{x}\right]=\mathrm{E}_{I_{\textsf{s},n},I_{\textsf{s},-n},I_{\textsf{m}}}\left[\mathrm{Pr}\left[X_{\textsf{s}_1}
\ge\theta_{\textsf{s}}(I_{\textsf{s},n}+I_{\textsf{s},-n}+I_{\textsf{m}})|\mathbf{X}=\mathbf{x}\right]\right]\nonumber\\
&\mathop=\limits^{(\mathrm{a})}&\underbrace{\mathrm{E}_{I_{\textsf{s},n}}\left[\exp\left(-\beta\theta_{\textsf{s}} I_{\textsf{s},n}\right)\right]}_{\triangleq\mathcal{L}_{I_{\textsf{s},n}}(z,\mathbf{x})|_{z=\beta\theta_{\textsf{s}}}} \underbrace{\mathrm{E}_{I_{\textsf{s},-n}}\left[\exp\left(-\beta\theta_{\textsf{s}} I_{\textsf{s},-n}\right)\right]}_{\triangleq\mathcal{L}_{I_{\textsf{s},-n}}(z,\mathbf{x})|_{z=\beta\theta_{\textsf{s}}}}\underbrace{\mathrm{E}_{I_{\textsf{m}}}\left[\exp\left(-\beta\theta_{\textsf{s}} I_{\textsf{m}}\right)\right]}_{\triangleq \mathcal{L}_{I_{\textsf{m}}}(z,\mathbf{x})|_{z=\beta\theta_{\textsf{s}}}},
\end{eqnarray}\setlength{\arraycolsep}{5pt}\endgroup}where (a) follows from $X_{\textsf{s}_1}\thicksim \exp\left(\beta\theta_{\textsf{s}}\right)$ \cite{CooperativeCachingandTransmissionDesigninClusterCentricSmallCellNetworks} and $\beta=P_{\textsf{s}}^{-1}/\left(\sum\nolimits_{k=1}^{K}x_k^{-\alpha_{\textsf{s}}}\right)$. To calculate $\psi_{\textsf{s}_1,\mathbf{X}}(T_n,\mathbf{x})$ according to (\ref{eqAppdPsis1}), we first calculate $\mathcal{L}_{I_{\textsf{s},n}}(z,\mathbf{x})$,  $\mathcal{L}_{I_{\textsf{s},-n}}(z,\mathbf{x})$ and $\mathcal{L}_{I_{\textsf{m}}}(z,\mathbf{x})$, respectively. Similar to (\ref{eqAppdPsimLIs}) and (\ref{eqAppdPsimLIm}), we have:
{\begingroup\makeatletter\def\f@size{10}\check@mathfonts
\def\maketag@@@#1{\hbox{\m@th\normalsize\normalfont#1}}\setlength{\arraycolsep}{0.0em}
\begin{eqnarray}
\mathcal{L}_{I_{\textsf{s},n}}(z,\mathbf{x})&=&\exp\left(-\pi\lambda_{\textsf{s}} T_nr_K^2\bigg({_2F_1}\bigg(-\frac{2}{\alpha_{\textsf{s}}},1;1-\frac{2}{\alpha_{\textsf{s}}};-\frac{z P_{\textsf{s}}} { x_K^{\alpha_{\textsf{s}}}}\bigg)-1\bigg)\right),\label{eqAppdPsis1LIsn}\\
\mathcal{L}_{I_{\textsf{s},-n}}(z,\mathbf{x})&=&\exp\left(-(1-T_n) {\frac{2\pi^2}{\alpha_{\textsf{s}}}\csc\left(\frac{2\pi}{\alpha_{\textsf{s}}}\right)}\lambda_{\textsf{s}}(z P_{\textsf{s}})^{2/\alpha_{\textsf{s}}}\right),\label{eqAppdPsis1LIsMIUSn}\\
\mathcal{L}_{I_{\textsf{m}}}(z,\mathbf{x}) &=&\exp\left(-{\frac{2\pi^2}{\alpha_{\textsf{m}}}\csc\left(\frac{2\pi}{\alpha_{\textsf{m}}}\right)\lambda_{\textsf{m}}}(z P_{\textsf{m}})^{2/\alpha_{\textsf{m}}}\right).\label{eqAppdPsis1LIm}
\end{eqnarray}
\setlength{\arraycolsep}{5pt}\endgroup}Substituting (\ref{eqAppdPsis1LIsn}), (\ref{eqAppdPsis1LIsMIUSn}) and (\ref{eqAppdPsis1LIm}) into (\ref{eqAppdPsis1}), we obtain $\psi_{\textsf{s}_1,\mathbf{X}}(T_n,\mathbf{x})$ as follows:
{\begingroup\makeatletter\def\f@size{10}\check@mathfonts
\def\maketag@@@#1{\hbox{\m@th\normalsize\normalfont#1}}\setlength{\arraycolsep}{0.0em}
\begin{eqnarray}\label{eqAppdPsis12}
\psi_{\textsf{s}_1,\mathbf{X}}(T_n,\mathbf{x})&=&\exp\left(-\pi\lambda_{\textsf{s}} T_nx_K^2\left({_2F_1}\left(-\frac{2}{\alpha_{\textsf{s}}},1;1-\frac{2}{\alpha_{\textsf{s}}};-\frac{\theta_{\textsf{s}} x_K^{-\alpha_{\textsf{s}}}}{\sum\nolimits_{k=1}^{K}x_k^{-\alpha_{\textsf{s}}}} \right)-1\right)\right)\nonumber\\
&&{\times}\:\exp\left(-{(1-T_n)\frac{2\pi^2}{\alpha_{\textsf{s}}}\csc\left(\frac{2\pi}{\alpha_{\textsf{s}}}\right)\lambda_{\textsf{s}}}\left( \frac{\theta_{\textsf{s}}}{\sum\nolimits_{k=1}^{K}x_k^{-\alpha_{\textsf{s}}}}\right)^{2/\alpha_{\textsf{s}}}\right)\nonumber\\
&&{\times}\:\exp\left(-{\frac{2\pi^2}{\alpha_{\textsf{m}}}{\csc\left(\frac{2\pi}{\alpha_{\textsf{m}}}\right)}\lambda_{\textsf{m}}} \left(\frac{\theta_{\textsf{s}} P_{\textsf{m}}}{P_{\textsf{s}}\sum\nolimits_{k=1}^{K}x_k^{-\alpha_{\textsf{s}}}} \right)^{2/\alpha_{\textsf{m}}}\right).
\end{eqnarray}\setlength{\arraycolsep}{5pt}\endgroup}Now, we calculate $\psi_{\textsf{s}_1}(T_n)$ by removing the condition of $\psi_{\textsf{s}_1,\mathbf{X}}(T_n,\mathbf{x})$ on $\mathbf{X}=\mathbf{x}$.  {Note that, the p.d.f. of $\mathbf{X}$ is given by
{\begingroup\makeatletter\def\f@size{10}\check@mathfonts
\def\maketag@@@#1{\hbox{\m@th\normalsize\normalfont#1}}\setlength{\arraycolsep}{0.0em}
\begin{eqnarray}\label{eqAppdPsis1pdf}
f_{{\mathbf{X}}}(\mathbf{x})=\begin{cases}
                               f_{X_K}(x_K), & \mbox{if } K=1, \\
                               f_{X_1,\cdots,X_{K-1}|X_K}(x_1,\cdots,x_{K-1}|x_K)f_{X_K}(x_K), & \mbox{if }K\ge2,
                             \end{cases}
\end{eqnarray}\setlength{\arraycolsep}{5pt}\endgroup}where $f_{X_K}(x_K)=\frac{2(\pi\lambda_{\textsf{s}}T_n)^K}{(K-1)!}x_K^{2K-1}e^{-\pi\lambda_{\textsf{s}}T_nx_K^2}$, $0<x_K<\infty$, is the p.d.f. of $X_K$ \cite{10}, and $f_{X_1,\cdots,X_{K-1}|X_K}(x_1,\cdots,x_{K-1}|x_K)=\prod\nolimits_{k=1}^{K-1}\frac{2x_k}{x_K^2}$, $0<x_k\le x_K$, is the conditional joint p.d.f. of $X_1,\cdots,X_{K-1}$, conditioned on $X_K=x_K$and is calculated by noting that given $X_K=x_K$, the $K-1$ SBSs are uniformly distributed in a circle of radius $x_K$ centered at $u_0$ \cite{DistanceDistributionsinFiniteUniformlyRandomNetworks}. Thus, by (\ref{eqAppdPsis12}) and (\ref{eqAppdPsis1pdf}), we~have:
{\begingroup\makeatletter\def\f@size{10}\check@mathfonts
\def\maketag@@@#1{\hbox{\m@th\normalsize\normalfont#1}}\setlength{\arraycolsep}{0.0em}
\begin{eqnarray*}\label{eqAppdPsis12finally}
\psi_{\textsf{s}_1}(T_n)&=&\begin{cases}
                    \displaystyle{\int_0^\infty} \psi_{\textsf{s}_1,\mathbf{X}}(T_n,\mathbf{x})f_{X_K}({x}_K)\mathrm{d}{x}_K, & \mbox{if } K=1, \\
                    \displaystyle{\int\nolimits_0^{x_K}\cdots\int\nolimits_0^{\infty}}\psi_{\textsf{s}_1,\mathbf{X}}(T_n,\mathbf{x}) f_{X_1,\cdots,X_{K-1}|X_K}(x_1,\cdots,x_{K-1}|x_K)f_{X_K}(x_K)\mathrm{d}x_1\cdots\mathrm{d}x_K, & \mbox{if }K\ge2.
                  \end{cases}
\end{eqnarray*}\setlength{\arraycolsep}{5pt}\endgroup}By using the changes of variables $u=\pi\lambda_{\textsf{s}}T_nx_K^2$, $t_k=\frac{x_k^2}{x_K^2}$, $k=1,\cdots,K-1$, and the definition of $B_{x,y}(\alpha_{x},\alpha_{y},T,\theta,u)$, we can get $\psi_{\textsf{s}_1}(T_n)$ in Theorem \ref{theorem1}.}
\section*{Appendix B: Proof of the Lemma \ref{lemmaOptimalityPropertiesp11}}
From (\ref{eqPsim}) and (\ref{eqPsiS1}), we know that $\psi_{\textsf{m}}>0$, $\psi_{\textsf{s}_1}(0)=0$ and $\psi_{\textsf{s}_1}(T_n)$ is an increasing function of $T_n$. Note that we consider the region $\psi_{\textsf{s}_1}(1)>\psi_{\textsf{m}}$. Thus, there exists a root $T_{\mathrm{th}}\in(0,1)$ such that $\psi_{\textsf{s}_1}(T_{\mathrm{th}})=\psi_{\textsf{m}}$. {Suppose the optimal solution  $\mathbf{T}^*$ satisfies $0<T_{n^*}^*\le T_{\mathrm{th}}$ for some $n^*\in\mathcal{N}$. Denote $\mathcal{N}^+\triangleq\{n\in\mathcal{N}|0< T_n^*\le 1\}$.\footnote{Note that, we have $\mathcal{N}^+\neq\emptyset$ due to the constraints in (\ref{Tncons1}) and (\ref{Tncons2}).} Note that $T_n^*=0$ for all $n\in\mathcal{N}\setminus\mathcal{N}^+$. Since $\sum\nolimits_{n\in\mathcal{N}}T_n^*=\sum\nolimits_{n\in\mathcal{N}^+}T_n^*=T_{n^*}^*+\sum\nolimits_{n\in\mathcal{N}^+\setminus\{n^*\}}T_n^*=M$, and $T_n^*\in(0,1]$ for all $n\in\mathcal{N}^+$, there exists $\epsilon_n\in[0,1)$ for all $n\in\mathcal{N}^+$ satisfying $\sum\nolimits_{n\in\mathcal{N}^+\setminus \{n^*\}}\epsilon_n=T_{n^*}^*$ and $\epsilon_n+T_n^*\in(0,1]$ for all $n\in\mathcal{N}^+\setminus \{n^*\}$. Since $T_{n^*}^*>0$, there exists $n^+\in\mathcal{N}^+\setminus\{n^*\}$ such that $\epsilon_{n^+}>0$. Now, we construct a feasible solution $\mathbf{T}^'$ to Problem \ref{problem1} by choosing $T_{n^*}^'=0$, $T_n^'=0$ for all $n\in\mathcal{N}\setminus\mathcal{N}^+$, and $T_{n}^'=T_n^*+\epsilon_n$ for all $n\in\mathcal{N}^+\setminus \{n^*\}$. Note that $T_{n^+}^'>T_{n^+}^*$, as $\epsilon_{n^+}>0$. Then, by the optimality of $\mathbf{T}^*$, we have:
{\begingroup\makeatletter\def\f@size{10}\check@mathfonts\def\maketag@@@#1{\hbox{\m@th\normalsize\normalfont#1}}\setlength{\arraycolsep}{0.0em}
\begin{eqnarray}\label{eqAppdpsiTdag}
\psi_{\mathrm{sch}_1}(\mathbf{T}^{'})-\psi_{\mathrm{sch}_1}(\mathbf{T}^{*})
&=&a_{n^*}(\psi_{\textsf{m}}-\psi_{\textsf{s}_1}(T_{n^*}^*))+\sum\nolimits_{n\in\mathcal{N}^+\setminus \{n^*\}}a_n(\psi_{\textsf{s}_1}(T_n^')-\psi_{\textsf{s}_1}(T_n^*))\le0.
\end{eqnarray}\setlength{\arraycolsep}{5pt}\endgroup}Since $0<T_{n^*}^*\le T_{\mathrm{th}}$ and $T_{n}^'=T_n^*+\epsilon_n\ge T_n^*$ for all $n\in\mathcal{N}^+\setminus \{n^*\}$, by the monotonicity of $\psi_{\textsf{s}_1}(x)$ w.r.t. $x$, we have $\psi_{\textsf{m}}-\psi_{\textsf{s}_1}(T_{n^*}^*)=\psi_{\textsf{s}_1}(T_{\mathrm{th}})-\psi_{\textsf{s}_1}(T_{n^*}^*)\ge0$ and $\psi_{\textsf{s}_1}(T_n^')-\psi_{\textsf{s}_1}(T_n^*)\ge0$. In addition, since $T_n^'\ge T_n^*$ for all $n\in\mathcal{N}\setminus\{n^*\}$ with strict inequality for at least $n^+$, we get $\sum\nolimits_{n\in\mathcal{N}^+\setminus \{n^*\}}a_n(\psi_{\textsf{s}_1}(T_n^')-\psi_{\textsf{s}_1}(T_n^*))>0$. Thus, we have $\psi_{\mathrm{sch}_1}(\mathbf{T}^{'})-\psi_{\mathrm{sch}_1}(\mathbf{T}^{*})>0$, which contradicts (\ref{eqAppdpsiTdag}). Therefore, by contradiction, we can prove that the optimal solution to Problem \ref{problem1}, i.e., $\mathbf{T}^*$, satisfies $T_n^*=0$ or $T_n^*\in(T_{\mathrm{th}},1]$ for all $n\in\mathcal{N}$, implying $\mathcal{N}^+=\{n\in\mathcal{N}|T_{\mathrm{th}}< T_n^*\le 1\}$ and $N_{\textsf{s}}^*\triangleq|\mathcal{N}^+|\in\left\{M,M+1,\cdots,\min\left\{\left\lceil\frac{M}{T_{\mathrm{th}}}\right\rceil-1,N\right\}\right\}$.

Suppose these exist $n_1\in\mathcal{N}^+$ and $\bar{n}_1\in\mathcal{N}\setminus\mathcal{N}^+$ (i.e., $T_{n_1}^*\in(T_{\mathrm{th}},1]$ and $T_{\bar{n}_1}^*=0$) such that $n_1>\bar{n}_1$ (i.e., $a_{n_1}< a_{\bar{n}_1}$). We construct a feasible solution $\mathbf{T}^{'}$ by choosing $T_{n_1}^'=T_{\bar{n}_1}^*$, $T_{\bar{n}_1}^'=T_{n_1}^*$ and $T_n^'=T_n^*$ for all $n\in\mathcal{N}\setminus\{n_1,\bar{n}_1\}$. Then, by the optimality of $\mathbf{T}^*$, we have:
{\begingroup\makeatletter\def\f@size{10}\check@mathfonts\def\maketag@@@#1{\hbox{\m@th\normalsize\normalfont#1}}\setlength{\arraycolsep}{0.0em}
\begin{eqnarray}\label{eqAppdpsiTdag2}
\psi_{\mathrm{sch}_1}(\mathbf{T}^{'})-\psi_{\mathrm{sch}_1}(\mathbf{T}^{*})
&=&(a_{n_1}-a_{\bar{n}_1})(\psi_{\textsf{m}}-\psi_{\textsf{s}_1}(T_{n_1}^*))\le0.
\end{eqnarray}\setlength{\arraycolsep}{5pt}\endgroup}Since $T_{n_1}^*\in(T_{\mathrm{th}},1]$, by the monotonicity of $\psi_{\textsf{s}_1}(x)$ w.r.t. $x$, we have $\psi_{\textsf{m}}-\psi_{\textsf{s}_1}(T_{n_1}^*)<\psi_{\textsf{m}}-\psi_{\textsf{s}_1}(T_{\mathrm{th}})=0$. In addition, by noting that $a_{n_1}- a_{\bar{n}_1}< 0$, we have $\psi_{\mathrm{sch}_1}(\mathbf{T}^{'})-\psi_{\mathrm{sch}_1}(\mathbf{T}^{*})>0$, which contradicts (\ref{eqAppdpsiTdag2}). Therefore, by contradiction, we prove that for all $n_1\in\mathcal{N}^+$ and $\bar{n}_1\in\mathcal{N}\setminus\mathcal{N}^+$, we have $n_1<\bar{n}_1$. That is, we have $\mathcal{N}^+=\{1,\cdots,N_{\textsf{s}}^*\}$ and $\mathcal{N}\setminus\mathcal{N}^+=\{N_{\textsf{s}}^*+1,N_{\textsf{s}}^*+2,\cdots,N\}$.

Consider $n_1,n_2\in\mathcal{N}^+$, $n_1<n_2$ (i.e., $a_{n_1}>a_{n_2}$). Suppose $T_{n_1}^*<T_{n_2}^*$. By the monotonicity of $\psi_{\textsf{s}_1}(x)$ w.r.t. $x$, we have $\psi_{\textsf{s}_1}(T_{n_1}^*)<\psi_{\textsf{s}_1}(T_{n_2}^*)$. Now, we construct a feasible solution $\mathbf{T}^'$ to Problem \ref{problem1} by choosing $T_{n_1}^'=T_{n_2}^*$, $T_{n_2}^'=T_{n_1}^*$, and $T_{n}^'=T_{n}^*$ for all $n\in\mathcal{N}\backslash\{n_1,n_2\}$. Thus, by the optimality of $\mathbf{T}^*$, we have
{\begingroup\makeatletter\def\f@size{10}\check@mathfonts\def\maketag@@@#1{\hbox{\m@th\normalsize\normalfont#1}}\setlength{\arraycolsep}{0.0em}
\begin{eqnarray}\label{eqAppdpsiTdag3}
\psi_{\mathrm{sch}_1}(\mathbf{T}^{'})-\psi_{\mathrm{sch}_1}(\mathbf{T}^{*})
&=&(a_{n_1}-a_{n_2})(\psi_{\textsf{s}_1}(T_{n_2}^*)-\psi_{\textsf{s}_1}(T_{n_1}^*))\le0.
\end{eqnarray}\setlength{\arraycolsep}{5pt}\endgroup}Since $a_{n_1}>a_{n_2}$ and $\psi_{\textsf{s}_1}(T_{n_1}^*)<\psi_{\textsf{s}_1}(T_{n_2}^*)$, we have $\psi_{\mathrm{sch}_1}(\mathbf{T}^{'})-\psi_{\mathrm{sch}_1}(\mathbf{T}^{*})>0$, which contradicts (\ref{eqAppdpsiTdag3}). Therefore, by contradiction, we prove that for any $n_1,n_2\in\mathcal{N}^+$, $n_1<n_2$, we have $1\ge T_{n_1}^*\ge T_{n_2}^*>T_{\mathrm{th}}$, we have $1\ge T_1^*\ge T_2^*\ge\cdots\ge T_{N_{\textsf{s}}^*}^*> T_{\mathrm{th}}$. By noting that $T_n^*=0$ for all $n\in\mathcal{N}\setminus\mathcal{N}^+$, implying $T_{N_{\textsf{s}}^*+1}=T_{N_{\textsf{s}}^*+2}=\cdots=T_N=0$. Therefore, we prove Lemma~\ref{lemmaOptimalityPropertiesp11}.}
\section*{Appendix C: Proof of the {Theorem} \ref{theorem2}}
To prove Theorem \ref{theorem2}, we rewrite (\ref{psi2Def}) as follows:
\begingroup\makeatletter\def\f@size{10}\check@mathfonts
\def\maketag@@@#1{\hbox{\m@th\normalsize\normalfont#1}}\setlength{\arraycolsep}{0.0em}
\begin{eqnarray}\label{eqApppsisch2}
\psi_{\mathrm{sch}_2}(\mathbf{T}) &=& \sum\limits_{n \in \mathcal{N}}{a_n}\left(\underbrace{\Pr \left[ \tau_{\textsf{m}} > \tau\right]}_{=\psi_{\textsf{m}}}\Pr[C_{2,n}=0]+ \sum\limits_{k=1}^{K}\underbrace{\Pr \left[ \tau_{\textsf{s}_2} > \tau |C_{2,n}=k\right]}_{\triangleq \psi_{\textsf{s}_2,k}}\Pr[C_{2,n}=k]\right),
\end{eqnarray}\setlength{\arraycolsep}{5pt}\endgroup where $\psi_{\textsf{m}}$ is already given by (\ref{eqPsim}) and $\Pr[C_{2,n}=k]=\binom{K}{k}T_n^k(1-T_n)^{K-k}$, $k=0,1,\cdots,K$. Thus, it remains to calculate $\psi_{\textsf{s}_2,k}$. Let $l_{\textsf{s},K}$ denote the $K$th nearest SBS in $\mathcal{C}_2$. We consider two cases, i.e., i) SBS $l_{\textsf{s},K}$ does not store file $n$, i.e., $l_{\textsf{s},K}\notin\mathcal{C}_{2,n}$ and  ii) SBS $l_{\textsf{s},K}$ stores file $n$, i.e., $l_{\textsf{s},K}\in\mathcal{C}_{2,n}$. Then, we have:
\begingroup\makeatletter\def\f@size{10}\check@mathfonts
\def\maketag@@@#1{\hbox{\m@th\normalsize\normalfont#1}}\setlength{\arraycolsep}{0.0em}
\begin{eqnarray}\label{eqpsis2Tn}
\psi_{\textsf{s}_2,k}&=&\Pr\left[W_{\textsf{s}}\log_2(1+\gamma_{\textsf{s}_2,n})>\tau,l_{\textsf{s},K}\notin\mathcal{C}_{2,n}| C_{2,n}=k\right]+\Pr\left[W_{\textsf{s}}\log_2(1+\gamma_{\textsf{s}_2,n})>\tau,l_{\textsf{s},K}\in\mathcal{C}_{2,n}| C_{2,n}=k\right]\nonumber\\
&=&\underbrace{\Pr\left[W_{\textsf{s}}\log_2\left(1+\frac{X_{\textsf{s}_2}}{I_{\textsf{s}}+I_{\textsf{m}}}\right)>\tau|l_{\textsf{s},K}\notin\mathcal{C}_{2,n},C_{2,n}=k\right] }_{\triangleq q_{k,1}} \Pr[l_{\textsf{s},K}\notin\mathcal{C}_{2,n}|C_{2,n}=k]\nonumber\\
&&{+}\: \underbrace{\Pr\left[W_{\textsf{s}}\log_2\left(1+\frac{X_{\textsf{s}_2}}{I_{\textsf{s}}+I_{\textsf{m}}}\right)>\tau|l_{\textsf{s},K}\in\mathcal{C}_{2,n}, C_{2,n}=k\right]}_{\triangleq q_{k,2}}\Pr[l_{\textsf{s},K}\in\mathcal{C}_{2,n}|C_{2,n}=k],
\end{eqnarray}\setlength{\arraycolsep}{5pt}\endgroup where {$X_{\textsf{s}_2}\triangleq{{{{\left| {\sum\nolimits_{{l_{\textsf{s}}} \in {{\cal C}_{2,n}}} {\sqrt {{P_{\textsf{s}}}} {h_{\textsf{s},l_{\textsf{s}}}}r_{\textsf{s},l_{\textsf{s}}}^{ - {\alpha_{\textsf{s}}}/2}} } \right|}^2}}}$}, $I_{\textsf{s}}\triangleq\sum \nolimits_{{l} \in {\Phi_{\textsf{s}}}\backslash {{\cal C}_{2}}}{P_{\textsf{s}}}{{\left| {{h_{\textsf{s},l}}} \right|}^2}r_{\textsf{s},l}^{ - {\alpha_{\textsf{s}}}}$ and $I_{\textsf{m}}\triangleq\sum\nolimits_{{l} \in {\Phi_{\textsf{m}}}} {P_{\textsf{m}}}{{\left| {{h_{\textsf{m},l}}} \right|}^2}\\r_{\textsf{m},l}^{ - {\alpha_{\textsf{m}}}}$. Note that, $\Pr[l_{\textsf{s},K}\notin\mathcal{C}_{2,n}|C_{2,n}=k] = 1-\frac{k}{K}$ and $\Pr[l_{\textsf{s},K}\in\mathcal{C}_{2,n}|C_{2,n}=k] = \frac{k}{K}$, $k=1,\cdots,K$. Thus, it remains to calculate $q_{k,1}$ and~$q_{k,2}$.

In the following, we focus on the calculation of $q_{k,1}$. Note that, $q_{k,2}$ can be calculated by following similar steps. We omit the details due to page limitation. When $k=K$, the case that $l_{\textsf{s},K}\notin\mathcal{C}_{2,n}$ and $\mathcal{C}_{2,n}=k$ cannot happen. In this case, we set $q_{k,1}=0$. Now, we calculate $q_{k,1}$ for the case that $l_{\textsf{s},K}\notin\mathcal{C}_{2,n}$ and $C_{2,n}=k$, $k=1,\cdots,K-1$. Let $X_1,\cdots,X_k$ denote the distances between the $k$ SBSs in $\mathcal{C}_{2,n}$ and $u_0$ and let $X_K$ denote the distance between the $K$th nearest SBS in $\mathcal{C}_2$ and $u_0$. Denote $\mathbf{X}\triangleq(X_1,\cdots,X_k,X_K)$. Further conditioning on $\mathbf{X}=\mathbf{x}$, we have:
{\begingroup\makeatletter\def\f@size{10}\check@mathfonts
\def\maketag@@@#1{\hbox{\m@th\normalsize\normalfont#1}}\setlength{\arraycolsep}{0.0em}
\begin{eqnarray}\label{eqAppdpsi1k}
q_{k,1,\mathbf{X}}(\mathbf{x})&\triangleq& \mathrm{Pr}\left[W_{\textsf{s}}\log_2\left(1+\frac{X_{\textsf{s}_2}}{I_{\textsf{s}}+I_{\textsf{m}}}\right)>\tau|\mathbf{X}=\mathbf{x},l_{\textsf{s},K}\notin\mathcal{C}_{2,n}, C_{2,n}=k\right]\nonumber\\
&=&\mathrm{E}_{I_{\textsf{s}},I_{\textsf{m}}}\left[\mathrm{Pr}\left[X_{\textsf{s}_2}
\ge\theta_{\textsf{s}}(I_{\textsf{s}}+I_{\textsf{m}})|\mathbf{X}=\mathbf{x},l_{\textsf{s},K}\notin\mathcal{C}_{2,n}, C_{2,n}=k\right]\right]\nonumber\\
&\mathop=\limits^{(a)}&\underbrace{\mathrm{E}_{I_{\textsf{s}}}\left[\exp\left(-P_{\textsf{s}}^{-1}/\left(\sum\nolimits_{i=1}^{k}x_i^{-\alpha_{\textsf{s}}}\right)\theta_{\textsf{s}} I_{\textsf{s}}\right)\right]}_{\triangleq\mathcal{L}_{I_{\textsf{s}}}(z,\mathbf{x})|_{z=P_{\textsf{s}}^{-1}/\left(\sum\nolimits_{i=1}^{k}x_i^{-\alpha_{\textsf{s}}}\right)\theta_{\textsf{s}}}}\underbrace{\mathrm{E}_{I_{\textsf{m}}}\left[\exp\left(-P_{\textsf{s}}^{-1}/\left(\sum\nolimits_{i=1}^{k}x_i^{-\alpha_{\textsf{s}}}\right)\theta_{\textsf{s}} I_{\textsf{m}}\right)\right]}_{\triangleq \mathcal{L}_{I_{\textsf{m}}}(z,\mathbf{x})|_{z=P_{\textsf{s}}^{-1}/\left(\sum\nolimits_{i=1}^{k}x_i^{-\alpha_{\textsf{s}}}\right)\theta_{\textsf{s}}}},
\end{eqnarray}\setlength{\arraycolsep}{5pt}\endgroup}where $\mathbf{x}\triangleq(x_1,\cdots,x_k,x_K)$, and (a) follows from that $X_{\textsf{s}_2}\thicksim \exp\left(P_{\textsf{s}}^{-1}/\left(\sum\nolimits_{i=1}^{k}x_i^{-\alpha_{\textsf{s}}}\right)\theta_{\textsf{s}}\right)$ \cite{CooperativeTransmissionviaCachingHelpers}. To calculate $q_{k,1,\mathbf{X}}(\mathbf{x})$ according to (\ref{eqAppdpsi1k}), we next calculate $\mathcal{L}_{I_{\textsf{s}}}(z,\mathbf{x})$ and $\mathcal{L}_{I_{\textsf{m}}}(z,\mathbf{x})$, respectively. Similar to (\ref{eqAppdPsimLIs}) and (\ref{eqAppdPsimLIm}), we have:
{\begingroup\makeatletter\def\f@size{10}\check@mathfonts
\def\maketag@@@#1{\hbox{\m@th\normalsize\normalfont#1}}\setlength{\arraycolsep}{0.0em}
\begin{eqnarray}
% \nonumber to remove numbering (before each equation)
  \mathcal{L}_{I_{\textsf{s}}}(z,\mathbf{x})&=&\exp\left(-\pi\lambda_{\textsf{s}}x_K^2 \left({_2F_1}\left(-\frac{2}{\alpha_{\textsf{s}}},1; 1-\frac{2}{\alpha_{\textsf{s}}},-\frac{z P_{\textsf{s}}}{x_K^{\alpha_{\textsf{s}}}}\right)-1\right)\right),\label{eqAppdPsi1kIs}\\
  \mathcal{L}_{I_{\textsf{m}}}(z,\mathbf{x})&=&\exp\left(-{\frac{2\pi^2}{\alpha_{\textsf{m}}} \csc\left(\frac{2\pi}{\alpha_{\textsf{m}}}\right)\lambda_{\textsf{m}}}(z P_{\textsf{m}})^{{2}/\alpha_{\textsf{m}}}\right).\label{eqAppdPsi1kIm}
\end{eqnarray}\setlength{\arraycolsep}{5pt}\endgroup}Substituting (\ref{eqAppdPsi1kIs}), (\ref{eqAppdPsi1kIm}) into (\ref{eqAppdpsi1k}), we have:
{\begingroup\makeatletter\def\f@size{10}\check@mathfonts
\def\maketag@@@#1{\hbox{\m@th\normalsize\normalfont#1}}\setlength{\arraycolsep}{0.0em}
\begin{eqnarray}\label{eqappqk1cond}
  q_{k,1,\mathbf{X}}(\mathbf{x})&=&\exp\left(-\pi\lambda_{\textsf{s}}x_K^2 \left({_2F_1}\left(-\frac{2}{\alpha_{\textsf{s}}},1; 1-\frac{2}{\alpha_{\textsf{s}}},-{\left(\sum\nolimits_{i=1}^{k}\left(\frac{x_K}{x_i}\right)^{\alpha_{\textsf{s}}}\right)}^{-1}\right)-1\right)\right)\nonumber\\
  &&{\times}\:\exp\left(-{\frac{2\pi^2}{\alpha_{\textsf{m}}} \csc\left(\frac{2\pi}{\alpha_{\textsf{m}}}\right)\lambda_{\textsf{m}}}\left(P_{\textsf{m}} P_{\textsf{s}}^{-1}/\left(\sum\nolimits_{i=1}^{k}x_i^{-\alpha_{\textsf{s}}}\right) \right)^{{2}/{\alpha_{\textsf{m}}}}\right).
\end{eqnarray}\setlength{\arraycolsep}{5pt}\endgroup}{Now, we calculate $q_{k,1}$ by removing the condition of $q_{k,1,\mathbf{X}}(\mathbf{x})$ on $\mathbf{X}=\mathbf{x}$. Note that, the p.d.f. of $\mathbf{X}$ is given by{\begingroup\makeatletter\def\f@size{10}\check@mathfonts
\def\maketag@@@#1{\hbox{\m@th\normalsize\normalfont#1}}\setlength{\arraycolsep}{0.0em}
\begin{eqnarray}\label{eqAppdPsi1kpdf}
  f_{\mathbf{X}}(\mathbf{x})=f_{X_1,\cdots,X_k|X_K}(x_1,\cdots,x_k)f_{X_K}(x_K),
\end{eqnarray}\setlength{\arraycolsep}{5pt}\endgroup}where $f_{X_K}(x_K)=\frac{2(\pi\lambda_{\textsf{s}})^K}{(K-1)!}x_K^{2K-1}e^{-\pi\lambda_{\textsf{s}}x_K^2}$, $0<x_K<\infty$,  is the p.d.f. of $X_K$ \cite{10}, and $f_{X_1,\cdots,X_k|X_K}\\(x_1,\cdots,x_k)=\prod_{i=1}^{k}\frac{2x_i}{x_K^2}$, $0<x_i\le x_K$, is the conditional p.d.f. of $X_1,\cdots,X_k$, conditioned on $X_K=x_K$ and is calculated by noting that given $X_K=x_K$, the $k$ SBSs are uniformly distributed in a circle of radius $x_K$ centered at $u_0$ \cite{DistanceDistributionsinFiniteUniformlyRandomNetworks}. Thus, by (\ref{eqappqk1cond}) and (\ref{eqAppdPsi1kpdf}), we have:
{\begingroup\makeatletter\def\f@size{10}\check@mathfonts
\def\maketag@@@#1{\hbox{\m@th\normalsize\normalfont#1}}\setlength{\arraycolsep}{0.0em}
\begin{eqnarray*}
  q_{k,1}&=&\int\nolimits_0^{x_K}\cdots \int\nolimits_0^{x_K} \int\nolimits_0^{\infty} q_{k,1,\mathbf{X}}(\mathbf{x})f_{\mathbf{X}}(\mathbf{x})\mathrm{d}x_1\cdots \mathrm{d}x_k\mathrm{d}x_K.
\end{eqnarray*}\setlength{\arraycolsep}{5pt}\endgroup}By using the changes of variables $u=\pi\lambda_{\textsf{s}}x_K^2$ and $t_i=\frac{x_i^2}{x_K^2}$, $i=1,\cdots,k$,  and the definition of $B_{x,y}(\alpha_{x},\alpha_{y},T,\theta,u)$ in (\ref{eqscheme1Bxy}), we can get $q_{k,1}$ in Theorem 2.}
\section*{Appendix D: Proof of the Lemma \ref{lemmaOptimalityPropertyeqproblem2Scheme2}}
The Lagrangian of the optimization in (\ref{eqnewpsi2T}) is given by
{\begingroup\makeatletter\def\f@size{10}\check@mathfonts
\def\maketag@@@#1{\hbox{\m@th\normalsize\normalfont#1}}\setlength{\arraycolsep}{0.0em}
\begin{eqnarray}
  L(\mathbf{T},\bm{\lambda},\bm{\eta},\nu)=\sum\nolimits_{n\in\mathcal{N}}a_n\psi_{\textsf{ms}}(T_n)+\sum\nolimits_{n\in\mathcal{N}}\lambda_nT_n +\sum\nolimits_{n\in\mathcal{N}}\eta_n(1-T_n)+\nu(M-\sum\nolimits_{n\in\mathcal{N}}T_n).
\end{eqnarray}\setlength{\arraycolsep}{5pt}\endgroup}where $\lambda_n\ge0$ and $\eta_n\ge0$ are the Lagrangian multipliers associated with (\ref{Tncons1}), $\nu$ is the Lagrangian multiplier associated with (\ref{Tncons2}), $\bm{\lambda}\triangleq(\lambda_n)_{n\in\mathcal{N}}$, and $\bm{\eta}\triangleq(\eta_n)_{n\in\mathcal{N}}$. Thus, we have:
{\begingroup\makeatletter\def\f@size{10}\check@mathfonts
\def\maketag@@@#1{\hbox{\m@th\normalsize\normalfont#1}}\setlength{\arraycolsep}{0.0em}
\begin{eqnarray}
  \frac{\partial L(\mathbf{T},\bm{\lambda},\bm{\eta},\nu)}{\partial T_n}=a_n\psi_{\textsf{ms}}^{'}(T_n)+\lambda_n-\eta_n-\nu.
\end{eqnarray}\setlength{\arraycolsep}{5pt}\endgroup}If $\mathbf{T}^*$ is an optimal solution of Problem \ref{problem2}, based on KKT conditions, i.e., (i) primal constraints: (\ref{Tncons1}), (\ref{Tncons2}), (ii) dual constraints: $\lambda_n\ge0$ and $\eta_n\ge0$ for all $n\in\mathcal{N}$, (iii) complementary slackness $\lambda_nT_n^*=0$ and $\eta_n(1-T_n^*)=0$ for all $n\in\mathcal{N}$, and (iv) $a_n\psi_{\textsf{ms}}^{'}(T_n^*)+\lambda_n-\eta_n-\nu=0$ for all $n\in\mathcal{N}$, we have: (a) if $T_n^*=0$, then $\lambda_n\ge0$, $\eta_n=0$, and $a_n\psi_{\textsf{ms}}^{'}(T_n^*)-\nu=-\lambda_n$, implying $a_n\psi_{\textsf{ms}}^{'}(T_n^*)\le\nu$; (b) if $T_n^*=1$, then $\lambda_n=0$, $\eta_n\ge0$, and $a_n\psi_{\textsf{ms}}^{'}(T_n^*)-\nu=\eta_n$, implying  $a_n\psi_{\textsf{ms}}^{'}(T_n^*)\ge\nu$; (c) if $0<T_n^*<1$, then $\lambda_n=0$, $\eta_n=0$, and $a_n\psi_{\textsf{ms}}^{'}(T_n^*)=\nu$. Therefore, we can prove (\ref{eqscheme2optprop}). In addition, by following similar steps as in the proof of Lemma~\ref{lemmaOptimalityPropertiesp11}, we can prove that $1 \ge T_1^*\ge T_2^*\ge\cdots\ge T_{N}^*\ge 0$. We omit the details due to page limitation.
% by themselves when using endfloat and the captionsoff option.
\ifCLASSOPTIONcaptionsoff
  \newpage
\fi

%\bibliographystyle{IEEEtran}
%\bibliography{Cooperative_caching_letter}

\bibliographystyle{IEEEtran}
%\bibliography{IEEEabrv,Cooperative_caching_letter}

\end{document}